\documentclass[12pt]{article}
\usepackage[reqno]{amsmath}
\usepackage{bbm}
\usepackage{epsfig}
\usepackage{array}
\usepackage{float}
\usepackage{a4}

\usepackage{a4wide}
\usepackage{wasysym}
\def\gs{\mathrel{
   \rlap{\raise 0.511ex \hbox{$>$}}{\lower 0.511ex \hbox{$\sim$}}}}
\def\ls{\mathrel{
   \rlap{\raise 0.511ex \hbox{$<$}}{\lower 0.511ex \hbox{$\sim$}}}}
\newcommand{\obb}{0\mbox{$\nu\beta\beta$}}
\newcommand{\onbb}{neutrinoless double beta decay }
\newcommand{\ba}{\begin{array}{c}}
\newcommand{\baz}{\begin{array}{cc}}
\newcommand{\bad}{\begin{array}{ccc}}
\newcommand{\bea}{\begin{equation} \begin{array}{c}}
\newcommand{\eea}{ \end{array} \end{equation}}
\newcommand{\ea}{\end{array}}

\newcommand{\dms}{\mbox{$\Delta m^2_{\odot}$}}
\newcommand{\dma}{\mbox{$\Delta m^2_{\rm A}$}}
\newcommand{\meff}{\mbox{$\langle m \rangle$}}
\newcommand{\eV}{\mbox{ eV}}

\newcommand{\be}{\begin{eqnarray}}
\newcommand{\ee}{\end{eqnarray}}

\newcommand{\ms}{\Delta m^2_{21}}
\newcommand{\ma}{\Delta m^2_{31}}
\newcommand{\sstwos}{\sin^2 2\theta_{12}}
\newcommand{\sss}{\sin^2 \theta_{12}}
\newcommand{\sch}{\sin^2 \theta_{13}}

\newcommand{\css}{\cos^2 \theta_{12}}
\newcommand{\csh}{\cos^2 \theta_{13}}
\def\ltap{\ \raisebox{-.4ex}{\rlap{$\sim$}} \raisebox{.4ex}{$<$}\ }

\newcommand{\meffnh}{\mbox{$\langle m \rangle$}^{\rm NH}}
\newcommand{\meffih}{\mbox{$\langle m \rangle$}^{\rm IH}}
\newcommand{\meffqd}{\mbox{$\langle m \rangle$}^{\rm QD}}
\newcommand{\meffnhmin}{\mbox{$\langle m \rangle$}^{\rm NH}_{\rm min}}
\newcommand{\meffihmin}{\mbox{$\langle m \rangle$}^{\rm IH}_{\rm min}}

\newcommand{\meffnhmax}{\mbox{$\langle m \rangle$}^{\rm NH}_{\rm max}}

\newcommand{\nme}{\mbox{nuclear matrix elements}}

\begin{document}

\title{\vspace{-2cm}
\hfill {\small OUTP--0505P}\\
\vspace{-0.3cm}
\hfill {\small TUM-HEP--591/05}\\
\vspace{-0.3cm}
\hfill {\small hep--ph/0506102} 
\vskip 0.5cm
\bf 
Neutrinoless Double Beta Decay and 
Future Neutrino Oscillation Precision Experiments}
\author{%\vskip -0.5cm
Sandhya Choubey$^a$\thanks{email: \tt sandhya@thphys.ox.ac.uk}~~~and~~
Werner Rodejohann$^b$\thanks{email: \tt werner$\_$rodejohann@ph.tum.de} 
\\\\
$^a${\normalsize \it Theoretical Physics, University of Oxford}\\
{\normalsize \it 1 Keble Road, Oxford OX1 3NP, 
UK}\\ \\
$^b${\normalsize \it Physik--Department, Technische Universit\"at M\"unchen,}\\
{\normalsize \it  James--Franck--Strasse, D--85748 Garching, Germany}
}
\date{}
\maketitle
\thispagestyle{empty}

\begin{abstract}

We discuss to what extent future precision measurements of neutrino 
mixing observables will influence the information we can draw from 
a measurement of (or an improved limit on) neutrinoless double beta decay. 
Whereas the $\Delta m^2$ corresponding to solar and atmospheric neutrino 
oscillations are expected to be known with good precision, 
the parameter $\theta_{12}$ will govern large part of the uncertainty. 
We focus in particular on the possibility of distinguishing the 
neutrino mass hierarchies and on setting a limit on the neutrino mass. 
We give the largest allowed values of the neutrino masses which allow 
to distinguish the normal from the inverted hierarchy. 
All aspects are discussed as a function of the uncertainty stemming from 
the involved nuclear matrix elements. 
The implications of a vanishing, or extremely small, effective mass are 
also investigated. 
By giving a large list of possible neutrino mass matrices and their 
predictions for the observables, 
we finally explore how a measurement of (or an improved limit on) 
neutrinoless double beta decay can help to identify the 
neutrino mass matrix if more precise values of the relevant 
parameters are known.

\end{abstract}
\newpage

\section{\label{sec:intro}Introduction}

The last few years have seen tremendous progress in our understanding
of the neutrino sector. The solar neutrino deficit is now known to 
be certainly due to neutrino flavor oscillations 
\cite{cl,ga,sksolar,sno,sno2,kl} with the best--fit oscillation 
parameters given as 
$\Delta m_\odot^2 \equiv \ms = 8.0\times 10^{-5}$ eV$^2$ and 
$\sin^2\theta_\odot \equiv \sss = 0.31$ \cite{sno2}. 
The atmospheric neutrino deficit can also be ascribed to 
flavor oscillations with a good degree of confidence \cite{skatm,k2k} 
with the best--fit oscillation parameters as 
$\Delta m_{\rm atm}^2 \equiv |\ma| = 2.1\times 10^{-3}$ eV$^2$ and 
$\sin^22\theta_{\rm atm} \equiv \sin^22\theta_{23} = 1$ 
\cite{skatm}. The upper limit on the third mixing angle 
$\theta_{13}$ is mainly determined by the 
reactor neutrino data \cite{chooz}, which when combined with the 
information obtained from the solar and atmospheric neutrino
experiments gives a bound of $\sch < 0.044$ at $3\sigma$
\cite{new,lisiglobal}.

Despite all these impressive achievements, 
there still remains a lot to be learned.
The arguably most fundamental 
question, namely whether neutrinos are Dirac or Majorana particles, 
remains still to be answered. 
If neutrinos are Majorana particles, we have nine physical parameters 
describing the $3\times 3$ light neutrino mass matrix and
determining them is one of the ultimate goals of neutrino physics. 
With the two $\Delta m^2$ and two mixing 
angles -- $\theta_{12}$ and $\theta_{23}$ --- known, there still 
remain five parameters to be tracked down. 
To be more precise, we still lack knowledge about the following points: 
\begin{itemize} 
\item Is there $CP$ violation in 
the lepton sector in analogy with that in the quark sector? 
\item Is the atmospheric neutrino mixing angle $\theta_{23}$  
exactly maximal? If not, does it lie above or below $\pi/4$? 
\item Is the third mixing angle $\theta_{13}$ exactly zero? 
\item How small is the absolute neutrino mass scale? 
\item What is the ordering of the neutrino masses, i.e., 
what is the neutrino mass hierarchy? 
\end{itemize} 
Answers to all or some of these question 
will certainly lead to better understanding of the underlying theory 
which gives rise to neutrino masses\footnote{
A rarely mentioned point is that for Majorana neutrinos 
one can, at least in principle, fully determine the complete mass matrix 
because it is symmetric. 
If the neutrinos are Dirac particles 
like quarks and charged leptons, their mass matrix is 
in general not symmetric and therefore parametrized by 
fifteen free parameters. However, since we only see left--handed 
weak currents, the  ``right--handed'' parts of the neutrino   
parameters are out of reach and hence not observable. 
}.\\ 

Neutrinoless double beta decay 
(\obb) is expected to be of crucial importance 
in answering some of the questions raised above. 
At the same time, answers to some the questions obtained 
from elsewhere will 
help in interpreting a positive or negative signal in the next 
generation \obb{} experiments. 
Goal of these experiments is the observation of the process 
\[ 
(A,Z) \rightarrow (A,Z-2) + 2 \, e^-~. 
\] 
The effective mass which 
will be extracted or bounded in a \obb{} experiment
is given by the following coherent sum: 
\be
\meff =  \left| \sum_i m_i \, U_{ei}^2 \right|~,
\label{eq:meff1}
\ee
where $m_i$ is the mass of the $i^{th}$ neutrino mass state, the sum 
is over all the light neutrino mass states and 
$U_{ei}$ are the matrix elements of the 
Pontecorvo--Maki--Nakagawa--Sakata (PMNS) neutrino mixing matrix \cite{PMNS}. 
We parametrize it as 
\be 
\label{eq:Upara}
U = \left( \bad 
c_{12} c_{13} & s_{12} c_{13} & s_{13}  \\[0.2cm] 
-s_{12} c_{23} - c_{12} s_{23} s_{13} e^{i \delta} 
& c_{12} c_{23} - s_{12} s_{23} s_{13} e^{i \delta} 
& s_{23} c_{13} e^{i \delta} \\[0.2cm] 
s_{12} s_{23} - c_{12} c_{23} s_{13} e^{i \delta} & 
- c_{12} s_{23} - s_{12} c_{23} s_{13} e^{i \delta} 
& c_{23} c_{13} e^{i \delta} \\ 
               \ea   \right) 
 {\rm diag}(1, e^{i \alpha}, e^{i \beta}) \, , 
\ee
where we have used the usual
notations $c_{ij} = \cos\theta_{ij}$, 
$s_{ij} = \sin\theta_{ij}$,
$\delta$ is the Dirac $CP$--violation phase, 
$\alpha$ and $\beta$ are the two Majorana 
$CP$--violation phases \cite{BHP80}.  
Thus we
can define $\meff$ in terms of the oscillation parameters, the Majorana 
phases and the neutrino mass scale. This means that \meff{} depends on 7 
out of 9 parameters contained in the neutrino mass matrix. 
In particular, the 
effective mass is a function of all unknowns of neutrino physics 
except for the Dirac phase and $\theta_{23}$. 
This means that a measurement of -- or a 
stringent limit on -- \obb{} will give us some information on the 
unknown parameters. Combined with 
complementary information from other 
independent types of experiments, this 
would reinforce our understanding of the 
neutrino sector.

There is a large 
list of available analyzes focusing on the connection of the known and 
unknown neutrino parameters with the effective mass 
\cite{ana1,PPR1,PPR2,nme_pha,cp,PPS}, 
for a recent review of the theoretical situation 
see \cite{rev_th}. 
In this work we want to focus on the mass ordering of the neutrinos,  
on the neutrino mass scale and on the implications of a very small 
or vanishing effective mass. 
The effective mass to be extracted from \onbb depends 
crucially on the neutrino mass spectrum. Of special 
interest are the following three extreme cases:  
\be 
\mbox{ normal hierarchy~(NH):} 
& |m_3| \simeq \sqrt{\dma} \gg |m_{2}| \simeq \sqrt{\dms} \gg |m_1|~,\\[0.3cm]
\mbox{ inverted hierarchy~(IH):} 
& |m_2| \simeq |m_1| \simeq \sqrt{\dma} \gg |m_{3}| ~,\\[0.3cm]
\mbox{ quasi--degeneracy~(QD):} 
& m_0 \equiv |m_3| \simeq |m_2| \simeq |m_1|  \gg \sqrt{\dma} ~.
\label{eq:mass}
\ee
The order of magnitude of the effective mass in those spectra is 
$\sqrt{\dms}, \sqrt{\dma}$ and $m_0$, respectively.\\ 

The best current limit on the effective mass is given by measurements of 
$^{76}$Ge established by the Heidelberg--Moscow collaboration \cite{HM} 
\be \label{eq:current} 
\meff \le 0.35\, z~{\rm eV}~, 
\ee
where $z={\cal O}(1)$ indicates that there is an uncertainty 
stemming from the nuclear physics involved in calculating the decay 
width of \obb. Similar results were obtained by the IGEX collaboration 
\cite{IGEX}. Several new experiments are currently running, under 
construction or in the planing phase. The NEMO3 \cite{NEMO} 
and CUORICINO \cite{CUORICINO} experiments are already taking data and reach 
sensitivities near the current limits. The next generation of experiments, 
with projects such as CUORE \cite{CUORE}, 
MAJORANA \cite{MAJORANA}, GERDA \cite{GERDA}, EXO \cite{EXO}, 
MOON \cite{MOON}, COBRA \cite{COBRA}, XMASS, DCBA \cite{DCBA}, 
CANDLES \cite{CANDLES}, CAMEO \cite{CAMEO}, (for a review see 
\cite{rev_ex}) 
will probe values of \meff{} one order of magnitude below the limit from 
Eq.\ (\ref{eq:current}).  
Thus we expect \meff{} to be probed down to 
$\simeq \sqrt{\dma} \simeq 0.04$ eV and 
it would be pertinent to ask if such a measurement could help 
us learn about the neutrino hierarchy\footnote{Of course those experiments 
aim also to put the controversial \cite{contr} evidence of part of the 
Heidelberg--Moscow collaboration to the test.}.\\

Maximal mixing in the solar neutrino sector is now disfavored at 
almost $6\sigma$ and this non--maximality of $\theta_{12}$ 
allows for the possibility to distinguish 
the NH from the IH case. This however depends on the 
value of the smallest neutrino mass state. We investigate this issue and 
give the value of the smallest neutrino mass state which allows 
to distinguish NH from IH. 
Another aspect, concerning the QD spectrum, is the 
size of the absolute neutrino mass scale.
We give and analyze a simple formula for a limit on $m_0$ in terms of 
$\theta_{12}, \theta_{13}$ and the limit (or value) of \meff. 
All these issues depend crucially on the uncertainties involved. 
In addition to the experimental errors, we would have to contend with
the theoretical uncertainties 
coming from unknown \nme{} and uncertainties in the 
values of the oscillation parameters.
As far the oscillation parameters are concerned,
the uncertainties are expected to be reduced sharply by 
the next generation experiments.
This will greatly 
reduce the expected range of \meff{} for a given mass spectrum. 
We give a detailed review of the improvements expected in each of the 
oscillation parameters and identify the parameters which would 
still lead to maximum uncertainty in the predicted value of \meff{} 
for a given mass spectrum. We take the uncertainties coming from 
our lack of knowledge of the \nme{} and explore the chances of 
determining the neutrino mass hierarchy from a positive \obb{} signal 
in the future. 
Furthermore, we analyze the consequences of a very small or 
even vanishing \meff{} if neutrinos are indeed Majorana particles.
Such a small \meff{} 
will influence the allowed range of values of 
the mixing parameters $\theta_{12}$, $\theta_{13}$, 
the Majorana phases 
and the smallest neutrino mass. 
We compare these implied range of parameter values with their 
current limits.
Finally, we perform a thorough scan of the phenomenologically 
viable neutrino mass models and give a list of possible 
neutrino mass matrices out of which the true one 
might be identified when future precision measurements  
of neutrino oscillation observables are 
combined with a measurement of or limit on \meff.\\

We begin by giving in Section \ref{sec:data} a detailed report of the 
existing range of values for the oscillation parameters. 
We list down the expected improvements in each of the parameters.
In Section \ref{sec:schemes} we probe the chances of distinguishing the 
IH from the NH from a future positive signal for \onbb$\!\!$. 
We take into account the current and future expected uncertainties 
on the oscillation parameters and include the role of the uncertainty 
in the \nme. In Section \ref{sec:m0}  we consider QD as the true mass spectrum 
of the neutrinos and study what limits one could set on the 
absolute/common neutrino mass scale $m_0$ from a measured 
value of $\meff$ and the current and future uncertainties on the 
oscillation parameters. Section \ref{sec:zero}  probes the situation one would 
have if we do not see any signal in the next set of \obb{} experiments
and yet believe that neutrinos are Majorana particles.
In Section \ref{sec:cor} we then list viable neutrino mass matrices and 
discuss how the number can drastically be reduced when measurements 
of \meff{} and the other oscillation parameters have been performed. 
We finally conclude in Section \ref{sec:concl}.

\section{\label{sec:data}Past, Present and Future data}
In this Section we shall present the status of the current global 
neutrino data and the prospects for its future improvement. First we 
give the current situation.

At the $3\sigma$ level, 
our current knowledge of the solar parameters within $3\sigma$
is limited to \cite{sno2,new}
\be
&7.0 \times 10^{-5} \eV^2 < \dms <  9.3\times 10^{-5} \eV^2~,& 
\label{eq:msrange}
\\
& 0.24 < \sss < 0.41~. &
\label{eq:sssrange}
\ee
The atmospheric mass squared difference and mixing 
angle at $3\sigma$ are known within 
\cite{skatm,lisiglobal}
\be
&1.3 \times 10^{-3} \eV^2 < \dma < 4.2 \times 10^{-3} \eV^2&,
\label{eq:marange}
\\
&0.33 < \sin^2\theta_{23} < 0.66&,
\ee
while the mixing angle $\theta_{13}$ at $3\sigma$ is restricted to lie 
below the value \cite{chooz,new,lisiglobal,kl2}
\be
\sch < 0.044~.
%0.051
\label{eq:schrange}
\ee

A very useful parameter related to the mass hierarchy of the neutrinos is 
the ratio of the solar and atmospheric mass squared differences, 
\be
R \equiv \frac{\dms}{\dma}~, \mbox{ with } 
0.017< R < 0.072~{\rm ~at~ 3\sigma}~. 
\label{eq:R}
\ee
Regarding the absolute value of the 
neutrino masses, several approaches exist. The most 
model--independent one is certainly the direct search for kinematical 
effects in the energy spectra of beta--decays. The Mainz \cite{mainz} 
and Troitsk \cite{troitsk} experiments gave upper limits on the 
electron neutrino mass of 2.3 eV at 95 $\%$ C.L.  
Cosmological observations imply typically more stringent limits, 
which however considerably depend on the data set 
and the priors used in the analysis. 
Combining the cosmic microwave radiation measurements by the WMAP 
satellite \cite{WMAP} with data on the large scale structure of the 
Universe and other data sets, gives limits on the sum of neutrino 
masses between 0.4 and 2 eV, see \cite{cosmo_dis} for a recent 
update of the situation.

Finally, there is currently no information on any of the three possible $CP$ 
phases, which take on values between zero and $2\pi$.\\

Let us now discuss the current uncertainties in the values 
of the parameters and 
their expected improvement.
We can see from Eqs.\ (\ref{eq:msrange})
and (\ref{eq:marange}) that we still have 
$53$\% ($14$\%) uncertainty\footnote{We define uncertainty of 
a quantity as the difference of the maximally and minimally allowed value 
divided by its sum and then multiplied by 100.} 
on the value of $\ma$ ($\ms$) at $3\sigma$. 
These ranges of allowed values 
are expected to reduce remarkably in the future: 
the uncertainty in $\ms$ is expected to reduce to $7$\% 
at the $3\sigma$ 
level from future measurement of reactor antineutrinos in KamLAND
\cite{kl2,th12new} and the value of $\ma$ is expected to be known with 
$\sim 5$\% at the $3\sigma$ level from the next generation accelerator
experiments involving superbeams in the next ten years \cite{huber10}. 

The uncertainty in the knowledge of $\sin^2\theta_{23}$ is at 
present $33$\% at $3\sigma$. 
This is expected to reduce to $\sim 20$\% 
in the next ten years after 
the results from the next generation superbeam experiments become
available \cite{huber10}.   
It is clear that the mixing angle $\theta_{23}$ plays no role 
for the study of neutrinoless double beta decay. Nevertheless, we 
shall see later on in Section \ref{sec:cor} that in order to 
identify the neutrino mass matrix it can be 
rather important to know ``how maximal'' $\theta_{23}$ actually is. 
Of particular importance in this respect 
is the question whether the 
atmospheric neutrino mixing is exactly maximal because this would 
point to the presence of an underlying symmetry. 
This deviation will be known to order $10\%$ after the 
next generation long--baseline experiments \cite{theta23}.
Comparative constraints are expected from future SK atmospheric 
neutrino data with statistics 50 times the current SK statistics
\cite{theta23_2}. 
Another way to state this is that we can identify if 
$\theta_{23} - \pi/4$ is of order $\sqrt{R}$ or smaller with 
next generation long--baseline experiments.  
A deviation from $\pi/4$ even smaller (i.e., order $R$) can be 
achieved by dedicated next generation experiments 
\cite{theta23,theta23_2,theta23_3} (see also \cite{maximality}).

The effective mass in neutrinoless double beta is 
crucially dependent on the values of the mixing parameters 
$\theta_{12}$ and $\theta_{13}$. The current uncertainty 
on the values of $\sss$ is $29$\% at the $3\sigma$ level
while $\sch$ is completely unknown. 
The uncertainty on the value of $\sss$ and the limit on $\sch$
is however expected to reduce in the future. While the conventional
beam experiments, MINOS, ICARUS and OPERA are expected to 
provide moderate improvement on $\sch$ \cite{huber10}, 
the Double--Chooz reactor 
experiment in France is expected to reduce the upper limit 
to $\sch \ltap 0.008$ at the $3\sigma$ level \cite{chooz2}. This upper 
limit could be improved further to $\sch \ltap 0.0025$ by 
the combination of the next generation 
beam experiments T2K \cite{t2k} and NO$\nu$A \cite{nova},
as well as by the second generation 
reactor experiments \cite{react13}. It should be 
borne in mind that any one of these experiments could 
even measure a non--zero $\theta_{13}$, if the true value 
of $\theta_{13}$ happens to fall within their range of sensitivity.

On the other hand, the improvement expected for 
$\sss$ from the currently running 
experiments is small \cite{th12new,th12}. The results 
from the third and the final Helium 
phase of the ongoing SNO experiment are expected to reduce the 
uncertainty on $\sss$ to not more than $\sim 21$\% \cite{kl2,skgd} 
and the KamLAND experiment 
is not expected to make any significant improvement on it 
\cite{th12new,th12,shika}. Doping the Super--Kamiokande (SK) detector 
with 0.1\% of Gadolinium \cite{gadzooks} could improve our knowledge 
on the true value of $\sss$ to $18$\% uncertainty \cite{skgd}.
The proposed/planned future experiments aiming to measure 
the very low energy $pp$ flux coming from the sun are also expected 
to give a better measurement of the solar mixing angle \cite{lownu}. 
However, 
even with very small experimental uncertainty of only 1\%, 
they are not expected to reduce the uncertainty on $\sss$ 
much better than $\sim 14$\% at $3\sigma$ \cite{th12new,roadmap}.
The only type of experiment that would provide extremely good 
measurement of the solar mixing angle is a long baseline reactor 
experiment with its baseline tuned to a Survival Probability 
minimum \cite{th12}, which is given by the condition 
$L_{min} \simeq 1.24 ~E/\rm MeV ~{\rm eV^2}/\ms$. 
This type of experiment could be used to measure the mixing angle 
$\sss$ down to $\sim 6$\% at $3\sigma$ \cite{th12new,minath12}.

The Dirac $CP$ phase $\delta$ has a faint chance of being measured 
in the next generation superbeam experiments \cite{t2k,nova}, provided
the true value of $\theta_{13}$ is not too small.
However, performing an unambiguous measurement, and 
especially establishing a signal for $CP$ violation, 
would probably need a beta beam facility 
or a neutrino factory \cite{nufac}. 
The two Majorana phases are  
measurable only in processes which violate lepton number. 
At present the only such process which seems to be viable 
experimentally is neutrinoless double beta decay \cite{DL2}. 
For detailed analyzes of how to extract information on Majorana phases 
from \obb{} we refer to \cite{PPR1,nme_pha,cp,PPS}. The bottom--line of 
this subject is that for given oscillation parameters 
the Majorana phases should not be too close to $0$ 
or $\pi/2$ and the uncertainty on both the experimental and theoretical 
side (i.e., the nuclear matrix elements) should be small \cite{PPR1,PPS}. 
In addition, 
the prospects of determining the Majorana $CP$ phases increase 
with increasing solar neutrino mixing angle. 

The quest for the limit on the absolute 
neutrino mass scale will witness attacks both via 
direct kinematical searches and cosmological measurements. 
The KATRIN \cite{KATRIN} experiment, 
currently under construction in Germany, 
is scheduled to start taking data in 2008 and 
is sensitive to neutrino masses down to 0.2 eV. Further cosmological 
probes will test the sum of neutrino masses down to 0.1 eV 
within this decade \cite{cosmo_dis}. 
Additional data sets and novel experiments can 
reduce this number by a factor of two \cite{cosmo_new}. 

Finally, the last piece of information needed to construct 
completely the neutrino mass matrix is the ordering of the neutrino
states -- the sign of $\ma$. 
If the sign was positive 
we would have a normal mass ordering, while for a negative sign  
we would have an inverted mass ordering.  
Typical approaches to identify 
sgn$(\ma)$ rely on using 
matter effects. Since the earth matter effect for modest baselines, 
and therefore modest matter densities, depends crucially on the 
size of the mixing angle $\theta_{13}$,
it becomes exceedingly difficult to study the mass 
hierarchy as $\theta_{13}$ becomes small. If $\theta_{13}$ was large, 
close to its current limit, there could be a small chance of 
measuring the mass hierarchy using the synergies between 
T2K and NO$\nu$A \cite{synergy}. 
However, the measurement would 
still not be very unambiguous and one would need either a beta beam 
facility or a neutrino factory for the hierarchy determination 
\cite{nufac}. Very large matter effects in the 
1--3 channel are expected for supernova neutrinos. Therefore, a 
supernova neutrino signal could in principle be used to determine the
sign of $\ma$ \cite{hie_sn}. Resonant matter effects in the 1--3 
channel are also 
encountered by atmospheric neutrinos as they cross large baselines 
in their passage through the earth. This can be exploited to 
probe the neutrino hierarchy both in water Cerenkov and large 
magnetized iron calorimeter detectors \cite{theta23_3,hie_atm}. 
Recently some 
novel ways of probing the mass hierarchy requiring very precise 
measurements and using the ``interference
terms'' between the different oscillation frequencies have been proposed
\cite{way_out_hie} (see also \cite{reac_hie}). 
However, it is understood that among all the 
neutrino parameters, the mass hierarchy determination 
is expected -- together with the determination of the Majorana phases --
to be the most challenging for the future experiments.
In this respect the \onbb is of some interest, since the 
question of distinguishing the neutrino mass hierarchy can be answered 
by \obb. This issue will be discussed in the following Section.

%%%%%%%%%%%%%%%%%%%%%%%%%%%%%%%%%%%%%%%%%%%%%%%%%%%%%%%%%%%%%%%%%%%%%%%
\section{\label{sec:schemes}Distinguishing the Neutrino Mass Schemes}
%%%%%%%%%%%%%%%%%%%%%%%%%%%

In this Section we present the phenomenology of \onbb 
in terms of the oscillation parameters 
$\ma$, $\ms$, $\sss$ and $\sch$. In particular we will look 
how easy or difficult it would be to distinguish the 
different neutrino mass schemes if we have a signal or 
a significantly improved limit for $\obb$.

For the NH scheme, for $m_1 \ll m_2 \ll m_3$ and therefore 
assuming that $m_1$ can be neglected, we have
\be
\meffnh \simeq \left| \sqrt{\ms} \, \sss \, \csh + \sqrt{\ma} \, \sch \, 
e^{2i(\beta - \alpha)} 
\right|
\label{eq:meffnh}~.
\ee
The maximal value, $\meffnhmax$, is achieved when both terms add up, 
which corresponds to $\alpha - \beta = \pm 2\pi, \pm \pi$ or 0. 
The minimal value of \meff{} corresponds to 
$\alpha - \beta = \pm \pi/2, 3\pi/2, 5\pi/2$ or $7\pi/2$, resulting
in (partial) cancellation 
between the two terms in Eq.\ (\ref{eq:meffnh}).
For non--zero values of $m_1$ between $10^{-2}$ and $10^{-3}$ eV 
there can be complete cancellation resulting in a vanishing effective mass. 
This will be subject of Section \ref{sec:zero}.

For the IH scheme, assuming that 
$m_3 \ll m_1 < m_2$ and neglecting $m_3$, we have
\be
\meffih \simeq \sqrt{|\ma|} \, \csh \, \sqrt{ 1 - \sstwos \, \sin^2
 \alpha} ~.
\label{eq:meffih}
\ee
Here the maximal (minimal) value is obtained when $\alpha = 0, \pi$ 
or $2\pi$ ($\alpha=\pi/2, 3\pi/2, 5\pi/2$ or $7 \pi/2$). 

Finally, for the QD mass spectrum
\be
\meffqd \simeq m_0 \left|  (\css + \sss \, e^{2i\alpha})\csh
+ e^{2i\beta} \, \sch \right|~.
\label{eq:meffqd}
\ee
If the terms proportional to $e^{2i\alpha}$ and $e^{2i\beta}$ add up 
(i.e., when $\alpha$ and $\beta$ take values of 
$ 0, \pi$ or $2\pi$) we have the maximal value of \meff, which is then just 
$m_0$.  
On the other hand, the minimal \meff{} is achieved when 
$\alpha$ and $\beta$ take values of 
$ \pi/2, 3\pi/2, 5\pi/2$ or $7\pi/2$. We will use this to set limits on 
$m_0$ in Section \ref{sec:m0}.

Hence, for a given mass scheme, 
the effective mass depends on
$\ma$, $\ms$, $\sss$, $\sch$ and on the Majorana phases (to be precise, 
on one of them, or on a combination thereof). 
In case of QD the absolute neutrino mass scale $m_0$ 
plays a decisive role as well. 
In the case of NH the smallest neutrino mass could be extremely 
important in deciding the degree of cancellation between 
the different terms, as discussed above.
The lack of knowledge of most of those 
parameters means that we can not give 
definite predictions for the value of $\meff$ for any of the 
three extreme neutrino mass schemes. 
We can however give a range of $\meff$ for NH, IH and QD, 
namely: 
%the three extreme mass schemes given by
\be
&0.0 < \meff < 0.007~{\rm eV}&,~~~~~{\rm NH} 
\\
&0.006~{\rm eV} < \meff < 0.065~{\rm eV}&,~~~~~{\rm IH} 
\label{eq:currlimih}
\\
&0.41~(0.07)~{\rm eV} < \meff < 2.3~(0.4)~{\rm eV}&,~~~~~{\rm QD} 
\ee
In order to obtain those values, we took the ranges of the oscillation 
parameters from Eqs.\ 
(\ref{eq:msrange},\ref{eq:sssrange},\ref{eq:marange},\ref{eq:schrange}) 
and inserted them in the expressions for the maximal 
and minimal values for \meff{} in the three extreme schemes under 
consideration. For the case of QD we used the limit from the direct 
kinematical search \cite{mainz,troitsk} and in brackets 
$m_0<0.4$ eV. 
Note that with the $3\sigma$ values the NH and IH cases slightly 
overlap. However, these ranges are expected to 
reduce remarkably in the future from more precise measurements of 
the solar and atmospheric neutrino parameters and from either a signal 
for a non--zero value of (or from better upper limits on) $\sch$. 
Nevertheless, already at the present stage the possibility of distinguishing 
the mass spectra from each other opens up. In particular, the case of 
deciding between NH and IH is the most interesting one. 
Distinguishing the QD spectrum from NH or IH will be at most a consistency 
check since the common neutrino mass of or above 0.2 eV will be probed 
by either direct searches or cosmology.

\begin{figure}[h]
\begin{center}
\epsfig{file=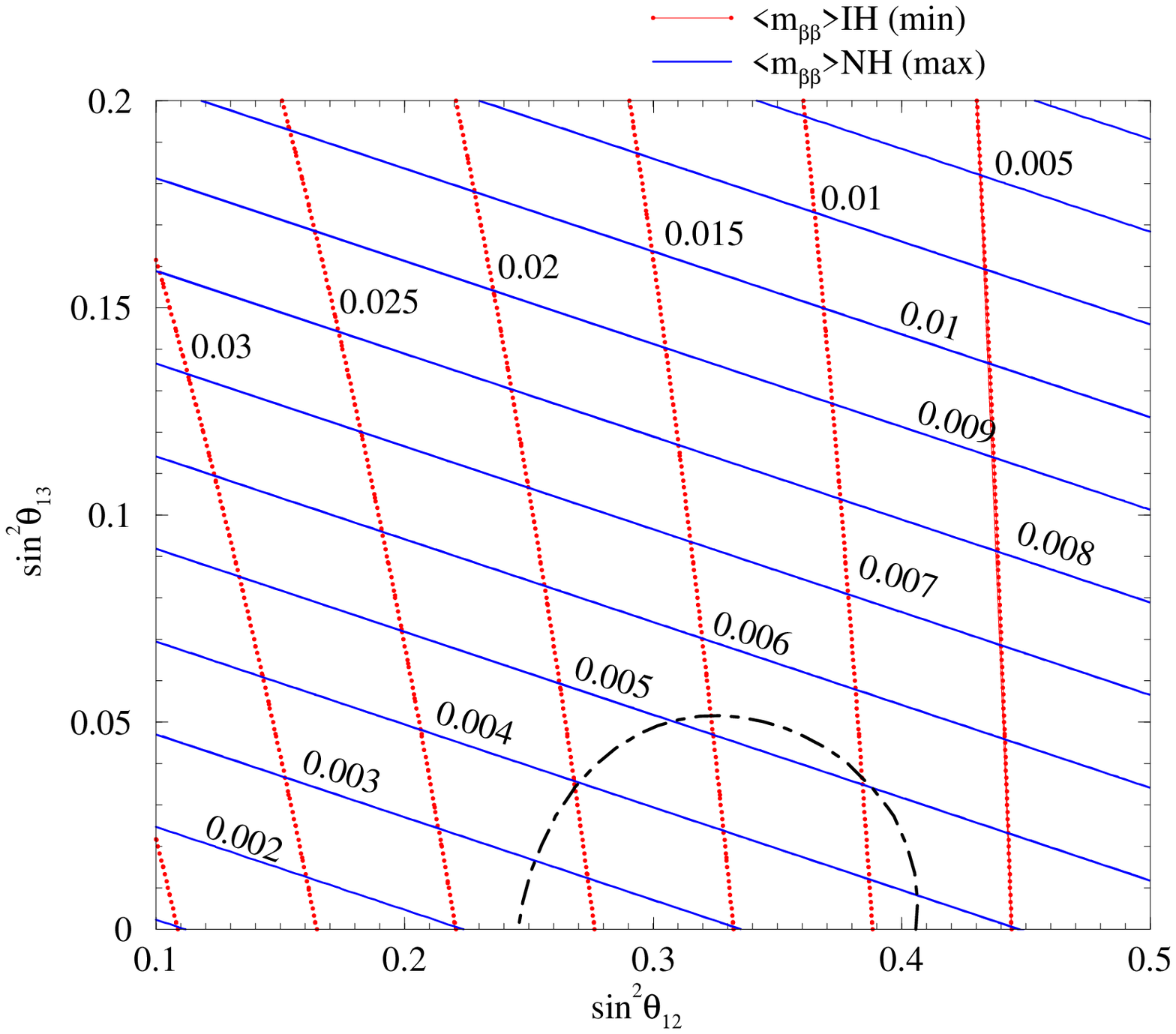,width=14cm,height=9.5cm}
\caption{\label{fig:th12-th13} 
Lines of constant maximum value of $\meff$ (in eV) 
for the NH scheme ($\meffnhmax$) and of 
constant minimum value of \meff{} for the IH scheme ($\meffihmin$). }
%\end{center}
%\end{figure}
%
%\begin{figure}[h]
%\begin{center}
\epsfig{file=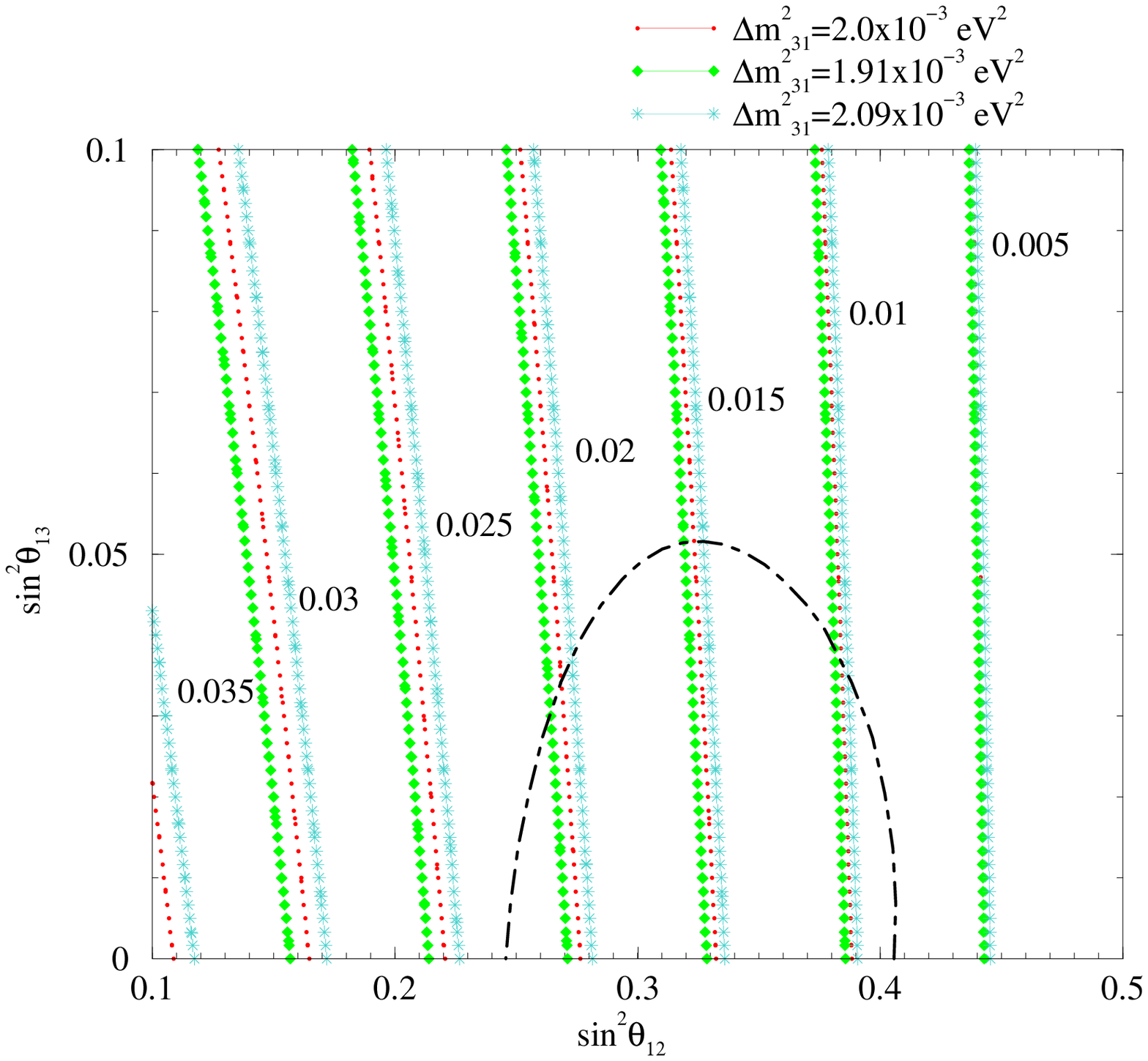,width=14cm,height=9.5cm}
\caption{\label{fig:th12-th13_IH_d31_future}
Lines of constant minimal value of \meff{} (in eV) in the IH case 
($\meffihmin$) for different $\Delta m^2_{31}$. }
\end{center}
\end{figure}

We present in Figure \ref{fig:th12-th13} the lines of 
constant maximal and minimal \meff{} for the cases of 
NH and IH, respectively, in the $\sss-\sch$ 
plane. 
We fixed the values of $\ma=0.002$ eV$^2$ and $\ms=8\times 10^{-5}$ eV$^2$. 
The dotted (red) lines show the 
lines of constant minimum value of \meff{} for the IH scheme 
($\meffihmin$) and the 
solid (blue) lines give the lines of constant maximum value of $\meff$
for the NH scheme ($\meffnhmax$). 
The Figure illustrates the well--known fact that $\meffihmin$ has a 
strong $\sss$ dependence, while $\meffnhmin$ depends on both 
$\sss$ and $\sch$. Also shown by the dot--dashed (black) line 
on the Figure are the current allowed 
values of $\sss$ and $\sch$ at the 3$\sigma$ level, obtained from 
a 2 parameter fit of the global oscillation data 
\cite{new}\footnote{The $3\sigma$ limit $\sch < 0.044$ 
given in the Introduction and 
Section \ref{sec:data} was obtained in a 1 parameter 
fit of the global neutrino data and is 
therefore different from the one in Figure \ref{fig:th12-th13}.}.  
As an example, one can see from the plot that if we know that 
the mass ordering is normal and assume that the smallest neutrino mass 
is negligible, values of the effective mass larger than roughly 0.0053 eV 
are incompatible with the currently allowed values of 
$\sss$ and $\sch$. The same would be the case when we know that the 
mass ordering is inverted, assume that the smallest neutrino mass 
is negligible and the effective mass is larger (or smaller) 
than 0.023 (0.008) eV.  

\begin{figure}[h]
\begin{center}
\includegraphics[width=14.0cm, height=9.5cm]
{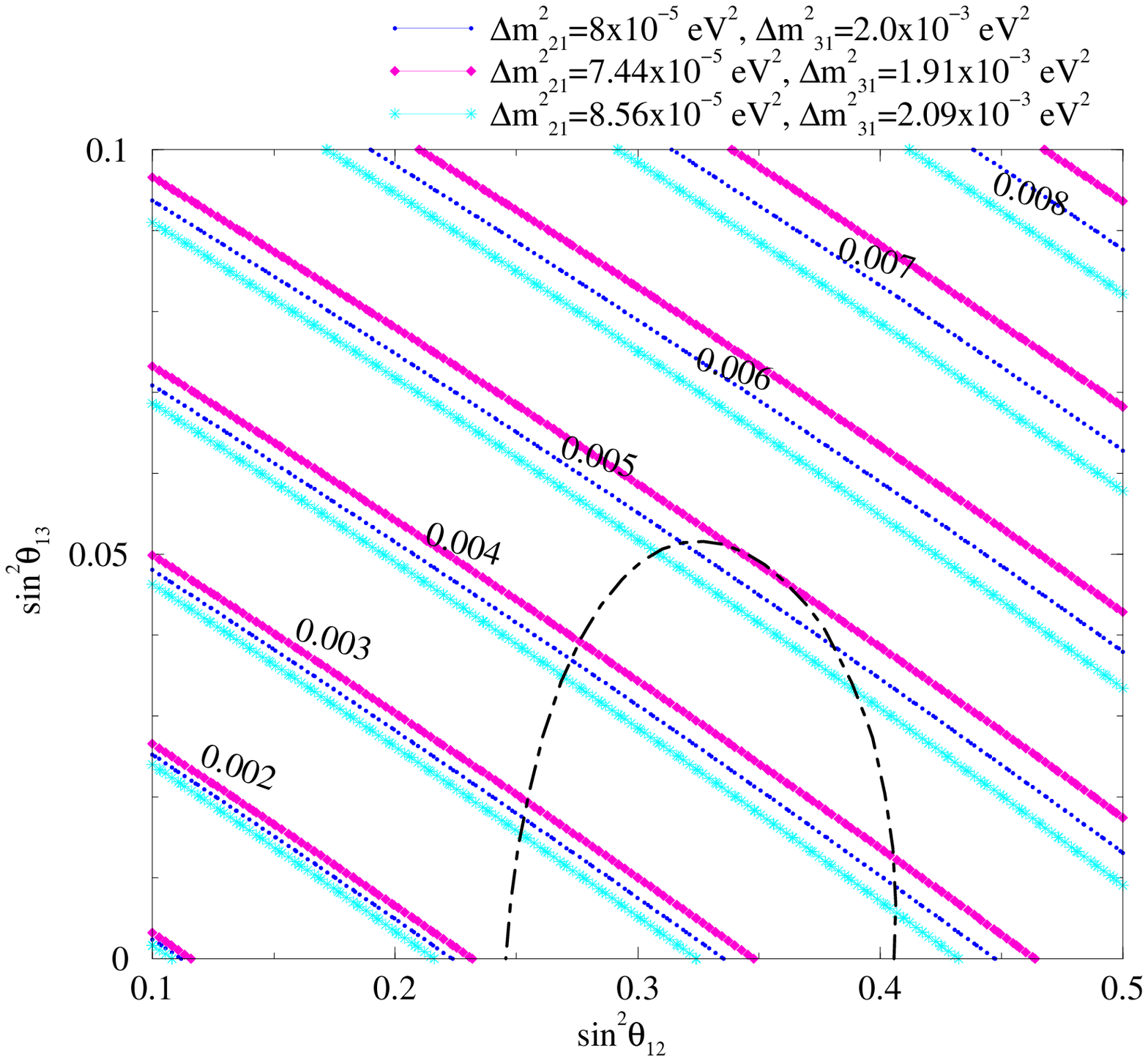}
\caption{Lines of constant maximal value of \meff{} (in eV) in the NH case 
($\meffnhmax$) for different $\Delta m^2_{21}$.}
\label{fig:th12-th13_NH_d21_future}
%\end{center}
%\end{figure}
%%%%%%%%%%%%%%%%%%%%
%%%%%%%%%%%%%%%%%%%%%%%%%%%%%%%%%%%%%%%%%
%\begin{figure}[t]
%\begin{center}
%\includegraphics[width=14.0cm, height=9.5cm]
%{diff_th12.eps}
%\caption{\label{fig:diff_th12} 
%The difference between the minimal value of \meff{} for IH 
%and the maximal value of \meff{} for NH, $\meffihmin - \meffnhmax$, 
%as a function of $\sin^2 \theta_{12}$.}

%\end{center}
%\end{figure}
%%%%%%%%%%%%%%%%%%%%
%%%%%%%%%%%%%%%%%%%%%%%%%%%%%%%%%%%%%%%%%
%\begin{figure}[t]
%\begin{center}
\includegraphics[width=15.5cm, height=9.5cm]
{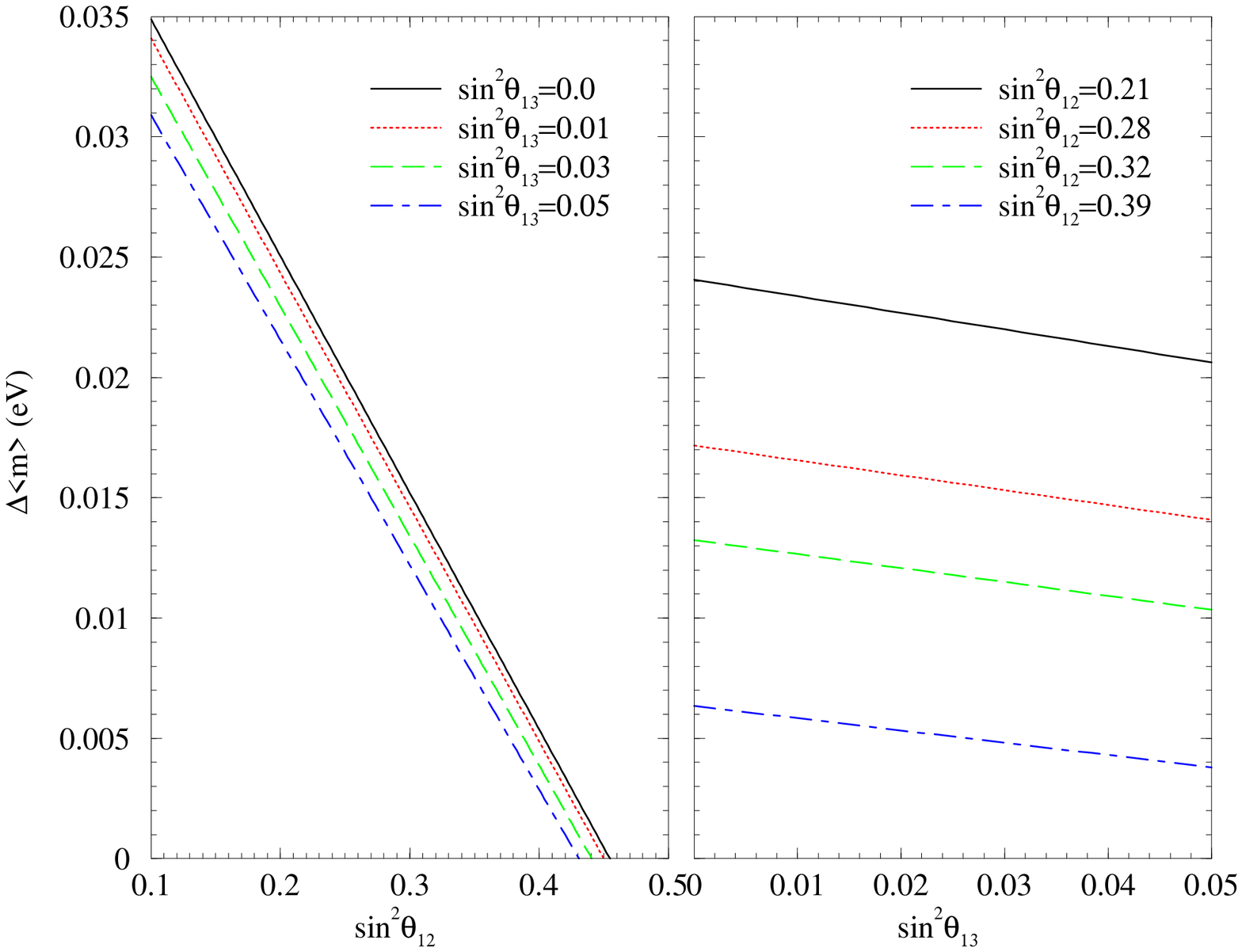}
\caption{\label{fig:diff}
The difference between the minimal value of \meff{} for IH 
and the maximal value of \meff{} for NH, $\meffihmin - \meffnhmax$, 
as a function of $\sin^2 \theta_{12}$ (left--hand panel) and 
$\sin^2 \theta_{13}$ (right--hand panel).
}
\end{center}
\end{figure}

%The maximal and minimal 
%values of $\meff$ plotted in Figure \ref{fig:th12-th13} are 
%for fixed values of $\ma=0.002$ eV$^2$ and $\ms=8\times 10^{-5}$ eV$^2$. 
We can see from Eqs.\ (\ref{eq:meffnh}) and (\ref{eq:meffih}) 
that the maximal and minimal values of \meff{} are crucially 
dependent on the values of the mass squared differences, especially on 
$\ma$. 
Since we are talking about a future measurement of the $\obb$ 
signal, we take the expected futuristic uncertainties mentioned in Section 
\ref{sec:data} on $\ma$ and $\ms$ to plot 
the iso--range of $\meffihmin$ and $\meffnhmax$ 
in Figures \ref{fig:th12-th13_IH_d31_future} and 
\ref{fig:th12-th13_NH_d21_future}. 
We can see clearly from the Figures that the uncertainty 
on the predicted value of $\meff$ due to the uncertainty in the 
value of $\ma$ and $\ms$ would become very small in the next 
ten years. Therefore, 
%unless otherwise stated, 
in what follows, we will keep $\ma$ and 
$\ms$ fixed at $\ma=0.002$ eV$^2$ and $\ms=8\times 10^{-5}$ eV$^2$.

Any future positive signal for $\obb$ will be able to distinguish
the IH scheme from the NH scheme if the measured value of 
$\meff$ would be such that it could be explained by the 
predicted $\meff$ for just one of the schemes and not the other.
In other words, if the difference between the predicted values 
for $\meff$ for the IH scheme and the NH scheme is larger than 
the error in the measured value of $\meff$, then it would be 
possible (assuming that the smallest neutrino mass is negligible) 
to experimentally find the neutrino mass hierarchy 
from a measurement of $\obb$. 
We therefore 
show the difference in the predicted value of $\meff$ for the two schemes,
\be \label{eq:Dme1}
\Delta\meff = \meffihmin - \meffnhmax~, 
\label{diff}
\ee
in left--hand panel of Figure \ref{fig:diff} as a function of 
$\sss$ for 4 fixed values of $\sch$ and in 
right--hand panel of Figure \ref{fig:diff} 
as a function of $\sch$ for 4 fixed values of $\sss$. 
We see that $\Delta\meff$ displays a strong dependence on $\sss$, whereas 
the dependence on $\sch$ is rather moderate. 
In fact, it holds that 
\be \label{eq:Dme2}
\Delta\meff = 
\frac{\sqrt{\dma} \, \cos^2 \theta_{13}}{1 + \tan^2 \theta_{12}}
\left( 
1 - \tan^2 \theta_{12} \, \left(1 + R + \tan^2 \theta_{13} \right) 
- \tan^2 \theta_{13}
\right) ~, 
\ee 
where $R$ is the ratio of the solar and atmospheric $\Delta m^2$ 
defined in Eq.\ (\ref{eq:R}).
Figure \ref{fig:diff} and Eq.\ (\ref{eq:Dme2}) demonstrate the well--known 
fact that distinguishing the normal and inverted hierarchy is easier 
for smaller values of 
$\sin^2 \theta_{12}$. The values of $\Delta\meff$, which are 
of the order 0.01 eV, represent a value of the maximal experimental 
uncertainty which an experiment should have in order to be able 
to distinguish NH from IH.\\

Up to here we neglected the complications which arise from the 
nuclear matrix elements (NME) involved in the calculation of the rate 
of \obb. There is neither a consensus in the literature on how large this 
uncertainty is, nor if there really is a large uncertainty at all 
\cite{simvog}. Let us therefore take a conservative point of view and 
analyze the situation as a function 
of the NME uncertainty.  If there is indeed no problem with the NME, 
the statements given up to this point apply.

\begin{figure}[t]
\begin{center}
\includegraphics[width=15.5cm, height=9.5cm]
{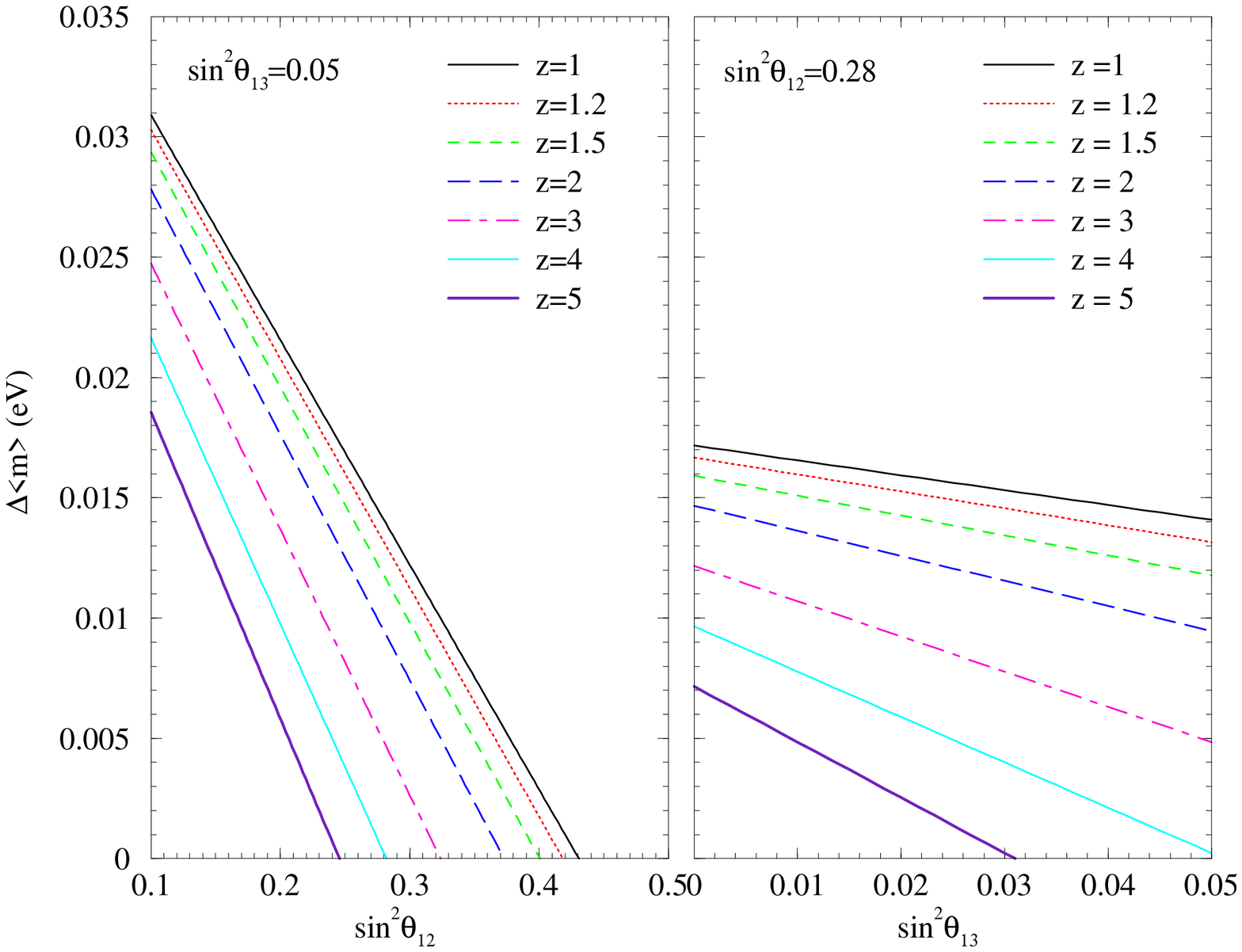}
\caption{The difference between the minimal value of \meff{} for IH 
and the maximal value of \meff{} for NH times the NME uncertainty, 
$\meffihmin - z \, \meffnhmax$, as a function of $\sin^2 \theta_{13}$ 
for different $z$ and $\sin^2 \theta_{12}=0.28$ (left--hand panel) and 
$\sin^2 \theta_{13}$ (right--hand panel).}
\label{fig:diff_nme}
%\end{center}
%\end{figure}
%%%%%%%%%%%%%%%%%%%%%%%%%%%%%%%%%%%%%%%%%%%%%%%
%\begin{figure}[t]
%\begin{center}
\includegraphics[width=17.0cm, height=9.5cm]
{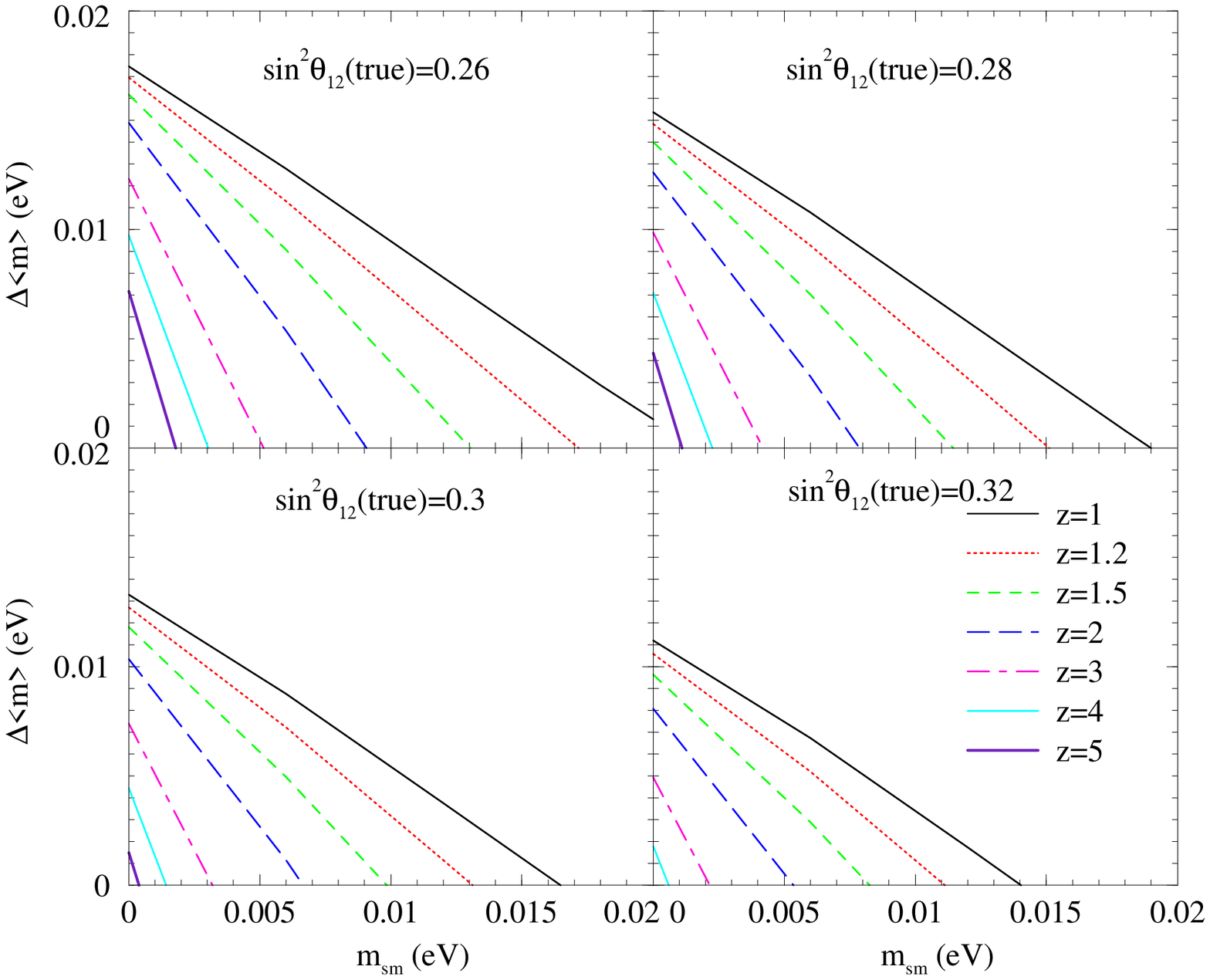}
\caption{\label{fig:m0_th12}
The difference between the minimal value of \meff{} for IH 
and the maximal value of \meff{} for NH, $\meffihmin - \meffnhmax$, 
as a function of the smallest neutrino mass $m_{\rm sm}$. The 4 panels 
are for 4 assumed ``true'' values of $\sss$. The values of 
$\sss$ are allowed to vary within 6\% of the assumed true value, 
while $\sch$ is allowed to vary between $0$ and $0.0025$.
}
\end{center}
\end{figure}

Given the possibility of large uncertainties on the values of the 
nuclear matrix elements, it is plausible to ask 
if one could possibly extract the true hierarchy of neutrino 
masses from a positive signal for $\obb$ in the future.
To that end, we 
%include the error coming from the lack of 
%knowledge of the \nme. We therefore 
look for the difference 
between $\meffihmin$ and $\meffnhmax$ after including the 
uncertainties coming from the \nme. 
A way to perform such an analysis has been developed in \cite{PPR1}, 
and here we follow a very similar approach:   
in order to parametrize the uncertainty coming from the \nme, we 
define the $\obb$ decay rate $\Gamma$ measured in any experiment such that 
\be
\sqrt{\Gamma} = x \, \meff~,
\ee
where $x$ contains all other factors involved in the $\obb$ 
decay, including the NME. Thus for any measured $\Gamma$ the 
measured value of $\meff$ would be
\be
\meff_{\rm measured} = \sqrt{\Gamma}/x~.
\ee
If $y$ is the smallest possible value for $x$, and if we assume that  
the uncertainty on $x$ originates solely  
from the uncertainty on the NME and is a factor of $z$ 
(with $z > 1$), then 
the range of $x$ is given by $y < x < z \, y$. Therefore, 
the uncertainty on 
the measured value of $\meff$ that we have due to the uncertainty 
on the NME is given by
\be
\Delta \meff_{\rm measured} = \frac{z-1}{z+1}~.
\label{eq:nmeuncert}
\ee
In order to be able to experimentally distinguish IH from NH with the 
help of a $\obb$ measurement, we must have 
$\meffihmin > \meffnhmax$ {\it after} including the 
uncertainty from the NME. Thus we need the condition
\be
\meffihmin - z \, \meffnhmax > 0
\ee
to ascertain the neutrino mass hierarchy. In the left and right panel of
Figure \ref{fig:diff_nme} 
we plot the difference $\meffihmin - z \, \meffnhmax$ predicted 
as a function of $\sss$ and $\sch$, respectively, including the 
uncertainty coming from the NME. We show the plots 
assuming a range of values for the NME uncertainty $z$. 
As a function of $\sss$ the quantity 
$\meffihmin - z \, \meffnhmax$ shows a similar behavior as for no 
uncertainty in the NME.  
The dependence on $\sch$ is rather weak for small $z$, but becomes 
larger with increasing $z$. 
From Figure \ref{fig:diff_nme} one can read of the maximal experimental 
uncertainty for a given $z$, $\sch$ and $\sss$ for which one could still 
be able to distinguish the mass hierarchy. 
For instance, taking 
$\sch$ close to its current limit and assuming 
$z = 2$ and $\sss = 0.3$, we have 
$\meffihmin - z \, \meffnhmax \simeq 0.01$ eV and the experimental 
uncertainty should not be larger than this value. For zero $\theta_{13}$ 
this value becomes larger by roughly $40\%$.\\

Up to now we considered the case of the smallest mass 
$m_{\rm sm}$ being zero, which 
would correspond to the neutrino mass matrix having a 
zero determinant, clearly 
a rather limited and special case.   
For a given set of oscillation parameters and the \nme{} uncertainty,  
it would be possible to distinguish between the NH and IH in a 
$0\nu\beta\beta$ experiment only if $m_{\rm sm}$ is less than a 
certain value. 
It is interesting to examine up to which values of the lightest 
neutrino mass one can distinguish NH from IH. 
Toward this exercise, 
we plot in Figure \ref{fig:m0_th12} 
the difference between the minimal value of \meff{} for IH 
and the maximal value of \meff{} for NH, $\meffihmin - \meffnhmax$, 
as a function of the smallest neutrino mass $m_{\rm sm}$. 
The NME uncertainty is also included in the analysis. 
For each of the 4 panels we assumed a certain 
``true'' value of $\sss$ with an optimistic uncertainty of 6\% 
\cite{th12new,minath12}. 
For $\sch$ we assumed that no positive signal is observed in any 
of the forthcoming experiments in the next ten years. 
Thus we allowed it to vary within $0$ and $0.0025$ \cite{huber10}. 
We see from the Figure that for $\sss = 0.3$ for example, 
when $z=1.5$ ($z=1$, {\it viz.} 
no NME uncertainty) and $\Delta \meff = 0.01$ eV, 
we can distinguish NH from IH if $m_{\rm sm}$ is 
smaller than 0.002 (0.004) eV. 
Of course, the value of $\Delta \meff$ obtained like this is again 
an upper limit on the experimental uncertainty which would still allow 
distinguishing NH from IH for a non--zero smallest mass. 
If $\sss = 0.26$ the situation looks more promising. For the same 
value of $\Delta \meff = 0.01$ eV we can distinguish NH from IH 
if $m_{\rm sm}$ is smaller than 0.005 (0.01) eV if $z=1.5$ ($z=1$). 
Other cases are easily read off Figure \ref{fig:m0_th12}. 

For a larger uncertainty in the value of $\sss$ and/or for a larger 
true value of $\sch$, the chances of distinguishing the IH from NH 
become worse and in addition only work for very small values of $m_{sm}$.

We see however that in principle values of $m_{\rm sm}$ up to $\sim 0.01$ eV 
are allowed in order to distinguish the normal from the inverted 
neutrino mass hierarchy.

\section{\label{sec:m0}Limit on the Neutrino Mass}
Another interesting aspect of \onbb is the 
possibility to set a limit on the absolute scale of the 
neutrino mass. The current upper value for the effective mass is somewhere 
between 0.3 and 1 eV, where this range of course 
has its origin in the NME uncertainty. The indicated values correspond to 
the QD mass spectrum, on which we wish to focus in this Section. 
As indicated in Section \ref{sec:schemes}, 
for a given common mass scale $m_0$, the lowest 
possible effective mass can be written as 
\be \label{eq:m0_meff}
\meff^{\rm QD}_{\min} = m_0 \, 
\left( 
|U_{e1}|^2 - |U_{e2}|^2 - |U_{e3}|^2 
\right) = \frac{1 - \tan^2 \theta_{12} - 2 \, |U_{e3}|^2 }
{1 + \tan^2 \theta_{12}}~. 
\ee
Hence, having a limit on the effective mass at hand, we can translate it 
into a limit on the neutrino mass. We write the experimental limit as 
\be \label{eq:zm0}
\meff^{\rm exp} = z \, \meff_{\min}^{\rm exp}~, 
\ee 
where $\meff_{\min}^{\rm exp}$ is defined as the limit on \meff{} 
obtained by using the largest available NME and $z \ge 1$ encodes 
again the NME uncertainty. Then, the limit on the neutrino mass reads 
\be \label{eq:m0_lim}
m_0 \le z \, \meff_{\min}^{\rm exp} \,  \frac{1 + \tan^2 \theta_{12}}
{1 - \tan^2 \theta_{12} - 2 \, |U_{e3}|^2 } 
\equiv z \, \meff_{\min}^{\rm exp} \,  f(\theta_{12}, \theta_{13})~.
\ee 
We have introduced here a function $f(\theta_{12}, \theta_{13})$ in this 
expression, which separates the information available from 
neutrino oscillation experiments from the 
information coming from \obb{} and $m_0$. 
We think that this formulation might be helpful in understanding how 
\onbb and the absolute neutrino mass scale are related. 
Currently the uncertainty on $f(\theta_{12}, \theta_{13})$ is 
around 50\%, $1.9 < f(\theta_{12}, \theta_{13})< 5.6$.  
It is expected to reduce to $\sim$ 21\%($\sim$ 9)~\% at $3\sigma$ 
if a low energy $pp$ solar neutrino experiment
(reactor experiment at the survival probability minimum) 
would be built. The uncertainty depends only little on the value of 
$\theta_{13}$. 
From the current limit on the effective mass, 
$\meff \le 0.35 \, z$ eV, with the accepted value of $z \simeq 3$ 
for $^{76}$Ge (see \cite{PPS}),  
and our current knowledge of 
$f(\theta_{12}, \theta_{13})$, we can set a limit on $m_0$ 
of 5.6 eV, clearly weaker than the limit from tritium experiments. 
Only for close to vanishing NME uncertainty (i.e., $z = 1$) we can 
reach values of $ m_0 \ls 2$ eV which are then comparable to ones from 
direct searches.

\begin{figure}[t]
\begin{center}
\includegraphics[width=14.0cm, height=9.5cm]
{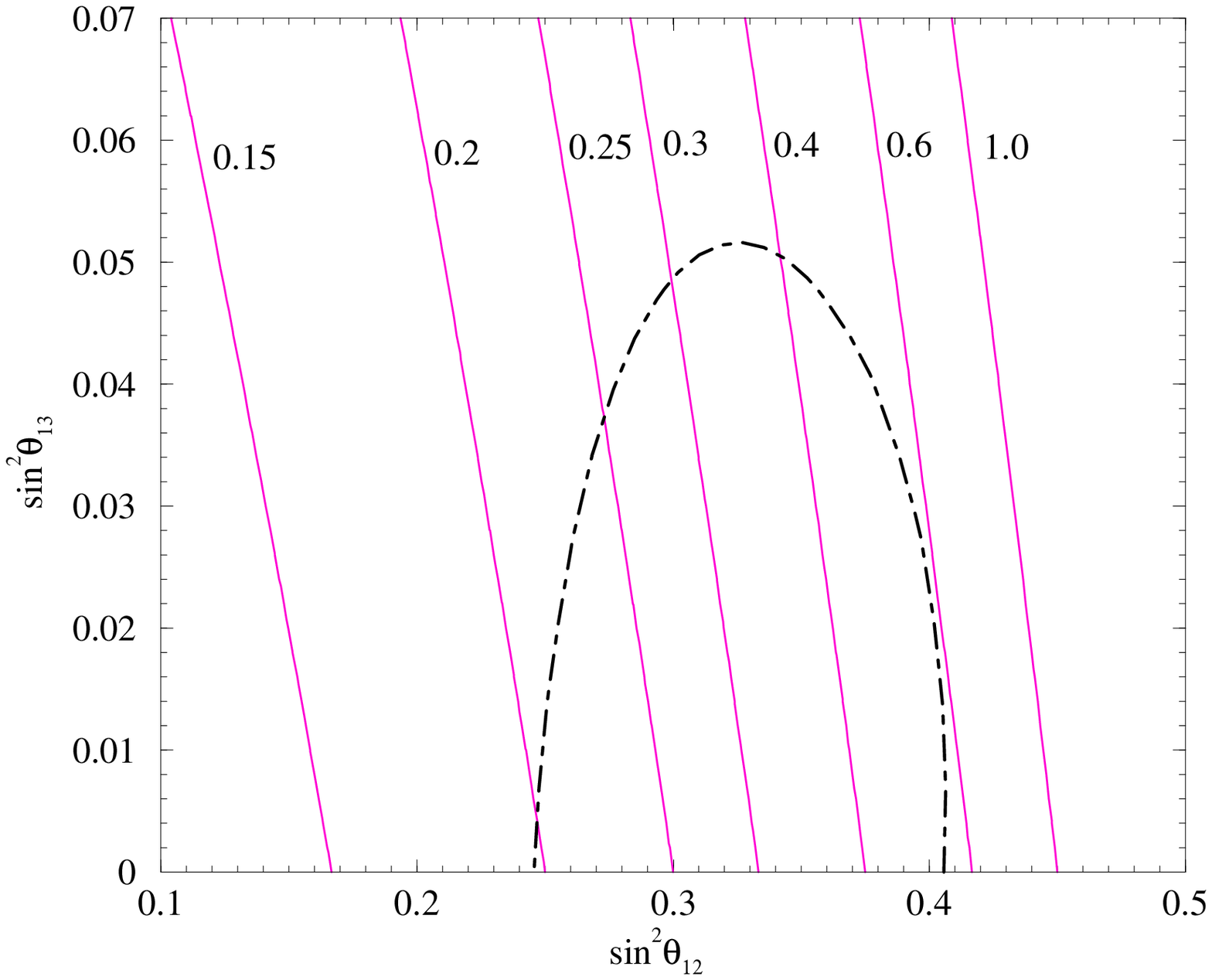}
\caption{\label{fig:limitm0}
The lines of constant $m_0$ in the 
$\sss-\sch$ plane, predicted 
for a QD mass spectrum. Also shown is the 
current $3\sigma$ allowed region in $\sss$ and $\sch$.
}
\end{center}
\end{figure}

In Figure \ref{fig:limitm0} we show the iso--contours of $m_0$ 
predicted for the QD mass spectrum (cf.\ Eq.\ (\ref{eq:m0_lim}))
in the $\sss-\sch$ plane. For each of the iso--contours we have 
assumed an illustrative value of $z \, \meff_{\min}^{\rm exp}=0.1$ eV. 
Also shown is the current 3$\sigma$ allowed 
area of $\sss$ and $\sch$. 
From the Figure we can see that for a measured value of 
$z \, \meff_{\min}^{\rm exp}=0.1$ eV, the constraint on the common mass 
scale for a QD spectrum would be: $0.2 ~{\rm eV} 
\ltap m_0 \ltap 0.6$ eV. For any other 
measured value of $z \, \meff_{\min}^{\rm exp}$ corresponding to a 
QD mass scheme, we can obtain the corresponding constraints on $m_0$ by 
scaling the above limit suitably, i.e., for 
$z \, \meff_{\min}^{\rm exp}=0.2$ eV we would have 
$0.4 ~{\rm eV}\ltap m_0 \ltap 1.2$ eV. 
The limits on $m_0$ given above are of course with our existing knowledge 
about $\theta_{12}$ and $\theta_{13}$. With reduction of the range 
of allowed values of $\theta_{12}$ and $\theta_{13}$, especially 
$\theta_{12}$, the constraints on 
$m_0$ are expected to reduce substantially, as discussed above. 
For instance, if $f(\theta_{12}, \theta_{13})$ was known with an 
uncertainty of $20\%$, say 
$2.7 < f(\theta_{12}, \theta_{13})< 4.0$, then we would have for 
$z \, \meff_{\min}^{\rm exp}=0.1~(0.2)$ eV that 
$0.3 ~{\rm eV} \ltap m_0 \ltap 0.4$ eV 
($0.6 ~{\rm eV} \ltap m_0 \ltap 0.8$ eV). 
Of course, if we have no signal for \obb, but just an upper limit 
on $z \, \meff_{\min}$, we have no longer an allowed range on $m_0$, 
but an upper limit corresponding to the largest value in the range. 
From the examples given above, one can note that 
for the QD mass spectrum, a measurement or 
a better constraint on \meff{} will 
lead to a stronger limit on the absolute neutrino mass 
scale compared to the current limit from 
direct kinematical searches.
 
%%%%%%%%%%%%%%%%%%%%%%%%%%%%%%%%%%%%%%%%%%%%%%%%%%%%%%%%%%%%%%%%%%%
\section{\label{sec:zero}Implications of a Vanishing effective Mass}
%%%%%%%%%%%%%%%%%%%%%%%%%%%%%%%%%%%%%%%%%%%%

It is well--known that not only a measurement but also 
a non--measurement of \onbb has significant influence on our 
knowledge of the unknown neutrino parameters. In this Section we wish 
to discuss a rarely studied subject, namely the impact 
that a very small upper limit on \meff{} 
would have\footnote{For a related earlier analysis, see \cite{xing}.}.  
An exactly vanishing \meff{} would 
correspond to a texture zero in the neutrino 
mass matrix in the charged lepton basis --
certainly an interesting feature. 
However, to prove/observe an exactly vanishing 
effective mass is a formidable task.

\begin{figure}[t]
\begin{center}
\includegraphics[width=14.0cm, height=9.5cm]
{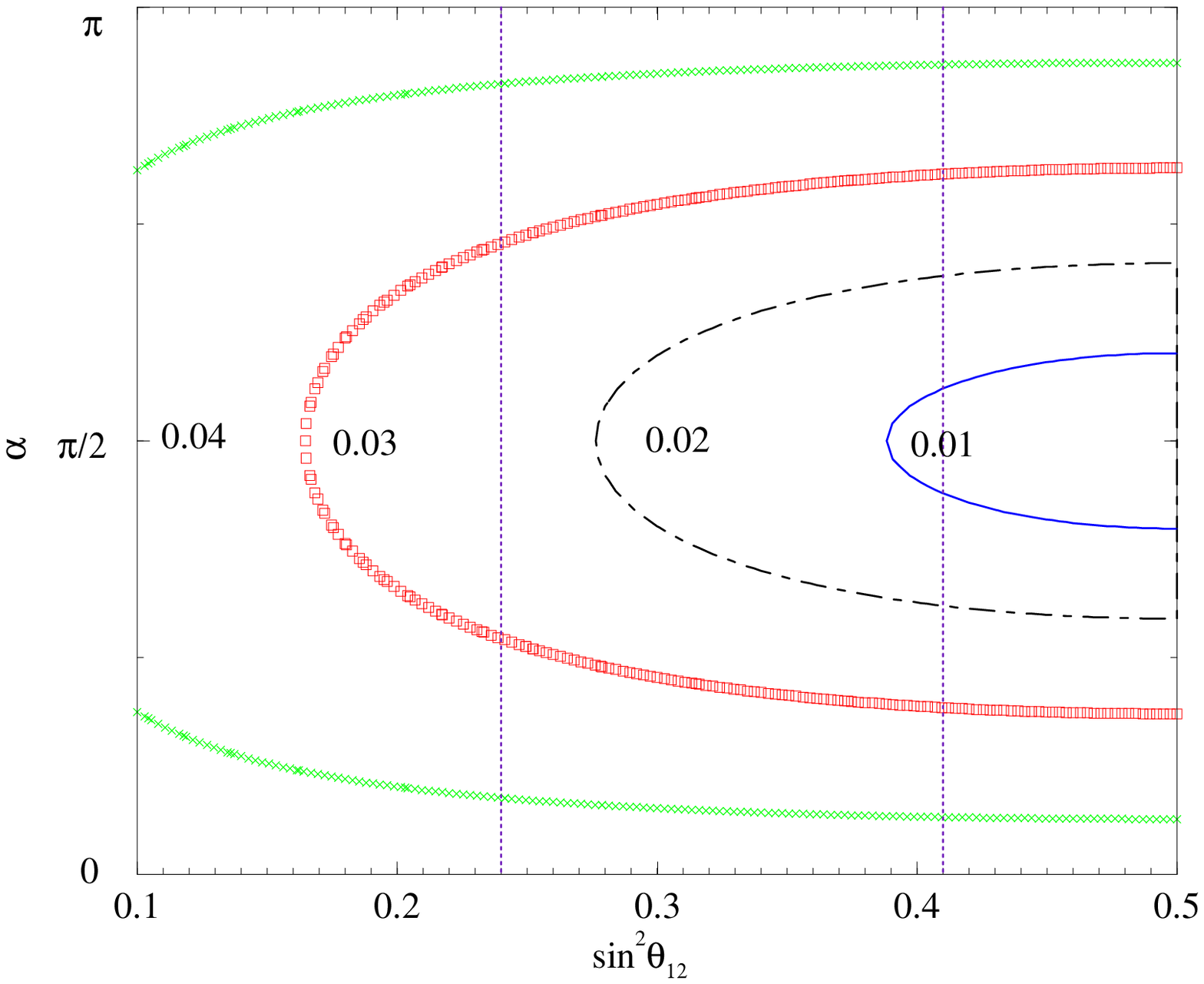}
\caption{\label{fig:zerom0_ih}Allowed values of $\sin^2 \theta_{12}$ and the 
Majorana phase $\alpha$ for a small upper limit on 
\meff{} (in eV) in case of 
IH. The dotted lines show the current $3\sigma$ limit on $\sss$. 
The allowed areas are to the right of the respective curves.
}
%\end{center}
%\end{figure}
%%%%%%%%%%%%%%%%%%%%%%%%%%%%%%%%%%%%%%%%%%%%%%%%%%%%%%%%%%%
%%%%%%%%%%%%%%%%%%%%
%\begin{figure}[t]
%\begin{center}
\includegraphics[width=14.0cm, height=9.5cm]
{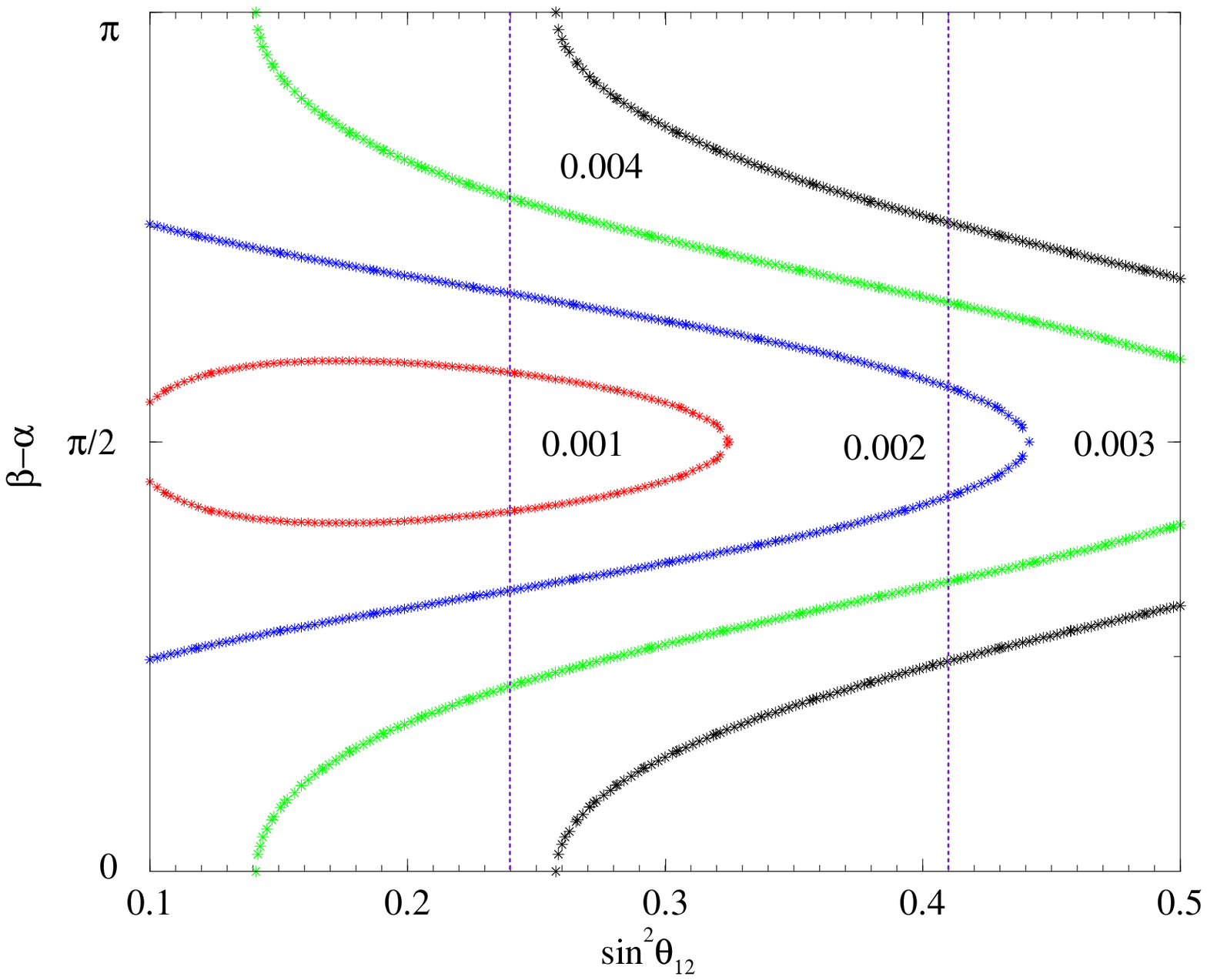}
\caption{\label{fig:zerom0_nh}Allowed values of 
$\sin^2 \theta_{12}$ and the Majorana phase combination $\beta - \alpha$  
for a small upper limit on \meff{} (in eV) in case of NH.
The dotted lines show the current $3\sigma$ limit on $\sss$.
The allowed areas are to the left of the respective curves.}
\end{center}
\end{figure}

In case of an inverted hierarchy we know that there is a lower limit 
of \meff{} which is --- when using current $3\sigma$ values of the 
oscillation parameters --- given by roughly 0.006 eV, see 
Eq.\ (\ref{eq:currlimih}).  
Let us assume that neutrinos are Majorana particles, 
have an inverted ordering and that the effective mass takes has an upper 
limit of order 0.01 eV. This is a situation which might arise
if we know from some other independent experiment
%long--baseline experiment 
that sgn$(\dma)=-1$ and the 
next generation \obb--experiments do not find a signal corresponding to 
a non--vanishing \meff, but give an upper limit on \meff{} still above the 
theoretical limit $\meff_{\rm min}^{\rm IH}$. 
%Less definite but also possible would be that 
%the latter holds and we just assume that sgn$(\dma)=-1$.  
Then we can infer from 
Eq.\ (\ref{eq:meffih}) the values of $\sin^2 \theta_{12}$ 
and the Majorana phase $\alpha$ which are still compatible with the data. 
In Fig.\ \ref{fig:zerom0_ih} we display the result for 
an experimental upper limit on \meff{} of 0.04, 0.03, 0.02, 0.01 eV, 
taking $m_3=0$, $\theta_{13}=0$ and $\ma = 0.002$ eV$^2$. 
We checked that the results are rather stable when we depart from these 
values within their current uncertainty. 
The current $3\sigma$ values of $\sin^2 \theta_{12}$ are also indicated 
in the Figure. 
Allowed is the area to the right of the respective curves. For instance, 
if $\meff < 0.02$ eV and $\sin^2 \theta_{12} = 0.3$ then $\alpha$ has to 
lie between 1.2 and 1.9, or $\alpha \simeq \pi/2 \pm 0.4$. 
Alternatively, if $\meff < 0.02$ eV and $\sin^2 \theta_{12} < 0.28$, then 
the IH case is ruled out.\\

Figure \ref{fig:zerom0_nh} shows a similar analysis for 
the case of NH with a smallest mass $m_1 = 0$. 
In our parametrization, 
as can be seen from Eq.\ (\ref{eq:meffnh}), it is 
the combination $\beta - \alpha$ which governs the 
destructive interference which leads to very small or zero \meff. 
We took very small upper limits on \meff{} of 0.004, 0.003, 0.002 and 0.001 eV 
(hence this discussion will not become realistic within the next 10 years) 
and chose $m_1=0$, $\sin^2 \theta_{13}=0.04$ 
$\ms = 8 \times 10^{-5}$ eV$^2$ 
and $\ma = 0.002$ eV$^2$. Note that for very small values of $\theta_{13}$, 
corresponding to $\sin^2 \theta_{13} \ll R\, \sin^2 \theta_{12} \sim 0.01$, 
the dependence on this combination of phases drops. 
In this part of the Figure it is the  
left of the respective curve which is allowed. 
For instance, for $\meff \ls$ 0.002 eV and a rather large value 
of $\sin^2 \theta_{12} = 0.4$ the phases have to be very close to $\pi/2$, 
namely $\beta - \alpha \simeq \pi/2 \pm 0.2$. 
Alternatively, if $\meff < 0.001$ eV and $\sin^2 \theta_{12} > 0.33$, 
then the NH case is ruled out. 

\begin{figure}[t]
 \begin{center}
\includegraphics[width=14.0cm, height=9.5cm]
{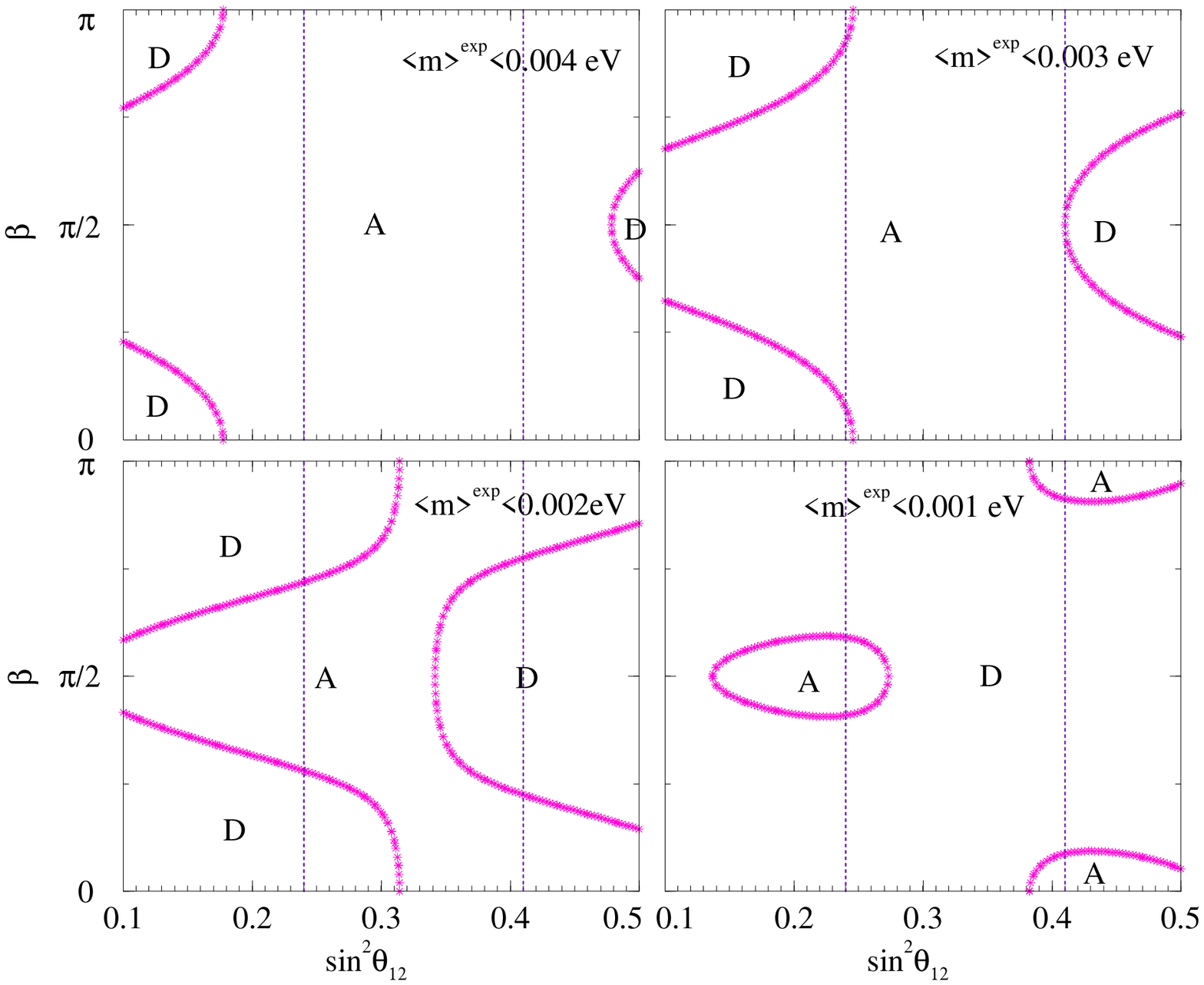}
\caption{\label{fig:m0_005}Allowed values of $\sin^2 \theta_{12}$ and 
the Majorana phase $\beta$ for a smallest neutrino mass of $m_1 = 0.005$ eV 
in case of NH for 4 different assumed upper limits on \meff{} and 
$\alpha=\pi/2$.
The dotted lines show the current $3\sigma$ limit on $\sss$. 
The allowed (disallowed) areas are marked with A (D).
}
%\end{center}
%\end{figure}
%%%%%%%%%%%%%%%%%%%%%%%%%%%%%%%%%%%%%%%%%%%%%%%%%%%%%%%%%%%
%%%%%%%%%%%%%%%%%%%%
%\begin{figure}[t]
%\begin{center}
\includegraphics[width=14.0cm, height=9.5cm]
{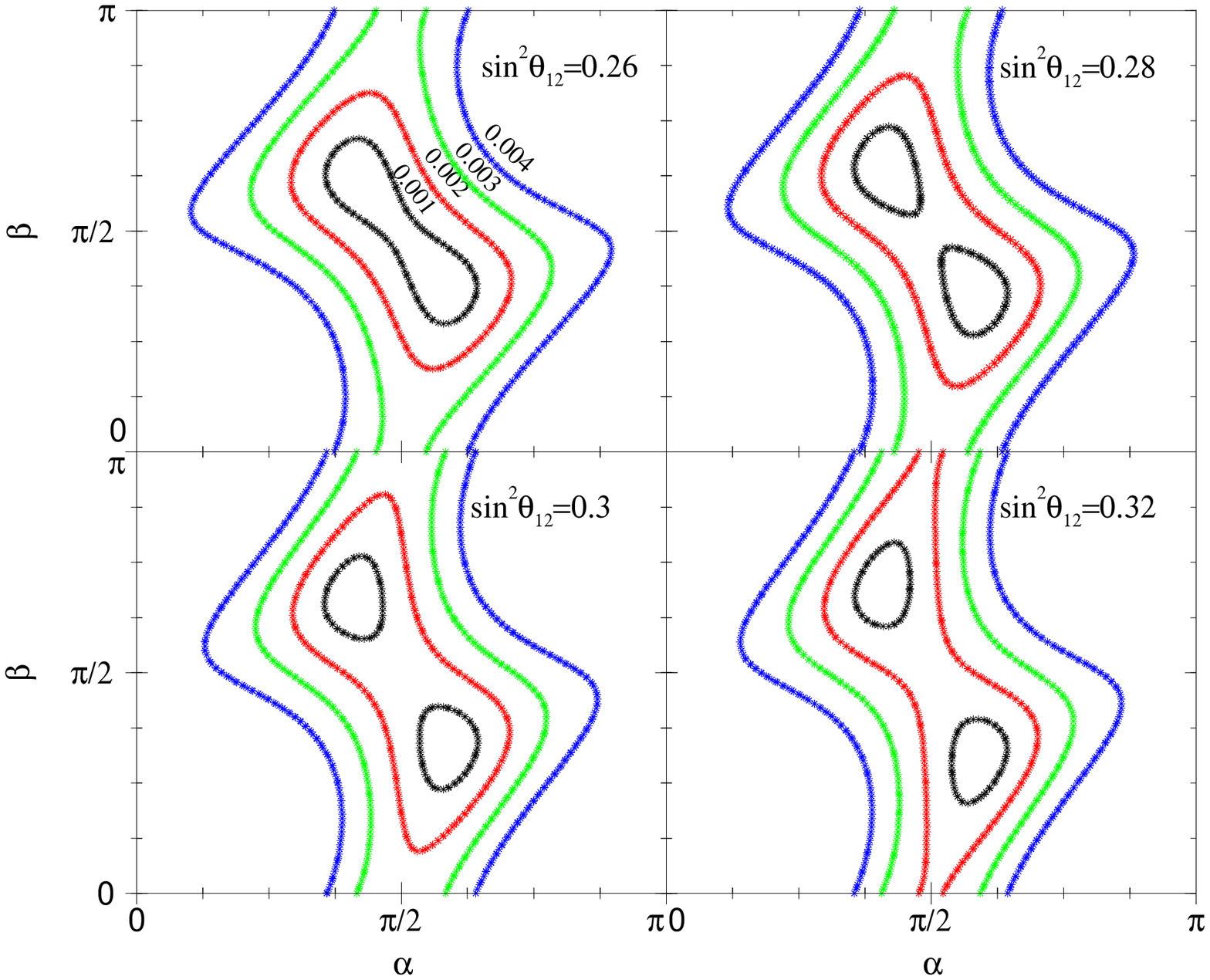}
\caption{\label{fig:m0_005phase}Allowed values of 
the Majorana phases $\alpha$ and $\beta$ for a smallest neutrino mass 
of $m_1 = 0.005$ eV 
in case of NH for 4 different assumed upper limits on \meff{}.
The 4 panels correspond to 4 different true values of $\sss$ shown 
on the Figure. Allowed are the areas within the respective curves.}
\end{center}
\end{figure}

As mentioned earlier in Section \ref{sec:schemes} we note here that 
in the case of NH, \meff{} could have a sizable dependence on the 
smallest neutrino mass state $m_1$. We plot therefore in Fig.\ 
\ref{fig:m0_005} the same as in Fig.\ 
\ref{fig:zerom0_nh} but for a smallest neutrino mass of $m_1 = 0.005$ eV. 
We took again $\sin^2 \theta_{13}=0.04$, 
$\ms = 8 \times 10^{-5}$ eV$^2$ 
and $\ma = 0.002$ eV$^2$ and chose the same four upper limits 
on \meff{} as above. 
With a non--vanishing $m_1$ Eq.\ (\ref{eq:meffnh}) is modified to 
\be
\meff \simeq \left| 
m_1 \, \cos^2 \theta_{12} \, \cos^2 \theta_{13} + 
\sqrt{m_1^2 + \dms} \, \sin ^2 \theta_{12} \, \cos^2 \theta_{13} \, 
e^{2 i \alpha} + \sqrt{\dma} \, \sin^2 \theta_{13} \, e^{2 i \beta} 
\right| ~ 
\label{eq:nhmeffm1}
\ee
and thus it is no longer $\beta - \alpha$ but both $\alpha$ and $\beta$ 
which play a role. In Figure \ref{fig:m0_005} we fixed $\alpha=\pi/2$
which gives a negative sign for the second term in 
Eq.\ (\ref{eq:nhmeffm1}).
We indicate the allowed (disallowed) values of the parameter
regions in the Figure with A~(D). 
If $\meff < 0.004$ eV and $\meff < 0.003$ eV the disallowed 
areas are already outside the currently allowed $3\sigma$ region 
of $\sin^2 \theta_{12}$ for $\alpha$ fixed at $\pi/2$. 
To show the dependence of our results on 
$\alpha$, we plot in 
Figure \ref{fig:m0_005phase} the allowed areas in 
the $\alpha$--$\beta$ plane for four different fixed values of 
$\sss$, keeping $\sch$ fixed at 0.04 and for $m_1=0.005$ eV. 
Allowed are the areas within the respective curves.

\begin{figure}[t]
\begin{center}
\includegraphics[width=14.0cm, height=9.5cm]
{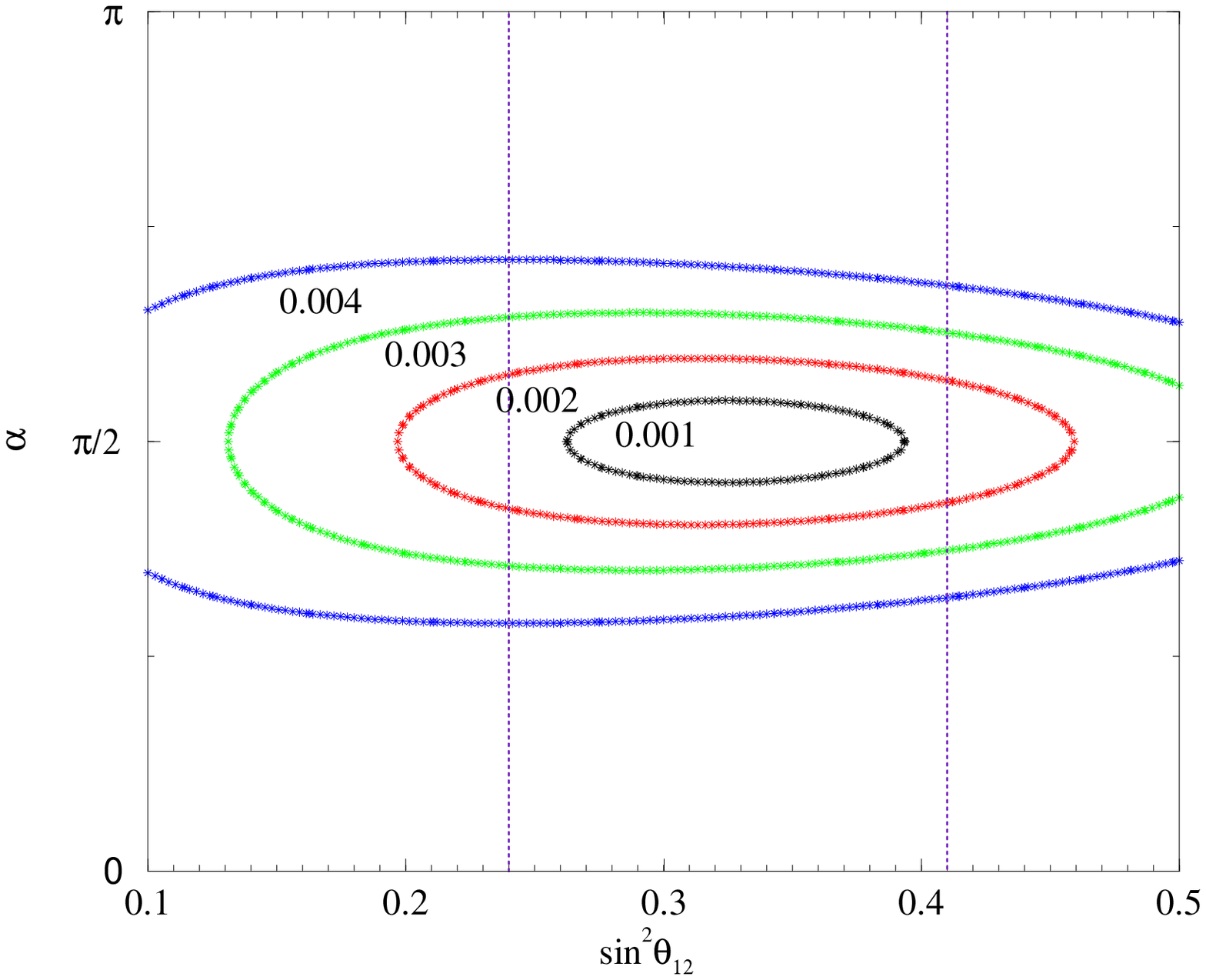}
\caption{\label{fig:m0_005th13}Allowed values of  
the Majorana phase $\alpha$ and $\sss$ for NH for 
$\sch=0$ and with
a non--zero smallest neutrino mass of $m_1 = 0.005$ eV 
in case of NH for 4 different assumed upper limits on \meff{}.
The dotted lines show the current $3\sigma$ limit on $\sss$. 
Allowed are the areas within the respective curves.}
\end{center}
\end{figure}

An interesting aspect is when $\theta_{13}$ vanishes but $m_1$ is non--zero. 
Then the effective mass is a function of the phase $\alpha$ alone: 
\be \label{eq:NHsp}
\meff \simeq \left| 
m_1 \, \cos^2 \theta_{12} + 
\sqrt{m_1^2 + \dms} \, \sin ^2 \theta_{12}  \, 
e^{2 i \alpha}  
\right| ~.
\ee
In Figure \ref{fig:m0_005th13} we show the areas in the 
$\sss$--$\alpha$ plane 
that would be allowed if we had an upper limit on 
\meff{} of 0.001 eV (black line), 0.002 eV (red line), 0.003 eV (green line)
and 0.004 eV (blue line). The areas outside the closed curves would 
be disallowed. We can see that for a limit of \meff=0.001 eV, all 
values of $\alpha$ could be possible while for \meff=0.004 eV, 
only the range $\pi/2 \pm 0.67$ is allowed. This 
allowed range of $\alpha$ is not expected to improve with the 
reduction of the uncertainty on $\sss$ since its $\theta_{12}$ 
dependence, as can be seen from the Figure, is very weak.

For values of the smallest mass of $m_1 \simeq 0.005$ eV and 
$\sin^2 \theta_{12} \simeq 0.3$ both terms in Eq.\ (\ref{eq:NHsp}) 
are of roughly 
the same magnitude (i.e., $m_1 \simeq m_2$) and we could write 
\[ 
\meff \simeq  m_{\rm NH} \, \sqrt{ 1 - \sstwos \, \sin^2 \alpha} 
~, \mbox{ where } m_{\rm NH} \equiv \sqrt{m_1^2 + \dms} \sim 0.01~\rm eV.
\] 
Note the similarity of this equation with Eq.\ (\ref{eq:meffih}) 
in case of IH. 
%There is however no known way to 
%probe such small values of $m_1$ and tell that $m_1 \simeq m_2$.  

To sum up, in certain cases it might be possible to significantly 
constrain the allowed values of the Majorana phases. Moreover, 
the allowed values of the solar neutrino oscillation parameter $\theta_{12}$ 
can be comparable to its current known range.

\section{\label{sec:cor}Neutrinoless Double Beta Decay, 
Future Neutrino Oscillation Data 
and the Identification of the Neutrino Mass Matrix} 

In this Section we wish to give a summary of typical 
neutrino mass matrices available in the vast literature. Predictions 
for and correlations between the neutrino observables are implied by each of 
the candidates and can be used to distinguish them. 
In particular, knowledge of the neutrino 
mass spectrum, $R$, $U_{e3}$, $\theta_{23}$ and of course the effective 
mass are then very helpful for identifying the correct mass matrix.  
A complete study of all 
possibilities is (and maybe can) not performed, 
however, we feel that the majority of the models 
will produce at the end of the day a mass matrix which will 
more or less correspond to one of the examples given here. 
For instance, there are 
many symmetries which eventually lead to a $\mu\tau$ symmetric mass 
matrix \cite{mutau}, see for instance Refs.\ \cite{mutau2} for 
some examples. 
For more models and details we refer to the excellent reviews 
available \cite{reviews}. 
We should remark that for the given examples, unless otherwise stated, the 
charged lepton mass matrix $m_\ell$ is diagonal, i.e., for the 
matrix $U_\ell$ diagonalizing $m_\ell m_\ell^\dagger$ holds that 
$U_\ell = \mathbbm{1}$, and therefore the PMNS matrix is just 
$U = U_\ell^\dagger \, U_\nu = U_\nu$, where $U_\nu$ diagonalizes 
$m_\nu$. Further note that most correlations focus on the quantities 
\meff, $R$, $\theta_{23}$, $|U_{e3}|$ or $\theta_{12}$. Interplay with the 
parameters describing $CP$ violation is rarely studied.\\

Let us focus first on the normal hierarchy, 
for which there 
is a very successful and well--known texture, namely \cite{NH1} 
\be \label{eq:mnuNH1}
m_\nu = m_0 \, 
\left(
\bad
a \, \epsilon^2 &  b \, \epsilon & d \, \epsilon \\[0.3cm]
\cdot & e & f \\[0.3cm]
\cdot & \cdot & g 
\ea
\right)~. 
\ee
Here $a,b,d,e,f,g$ are complex parameters of order one and $\epsilon$ 
is a real and small parameter typically of order of the 
Cabibbo angle $\lambda \simeq 0.22$. 
Such a mass matrix preserves at leading order the lepton charge $L_e$ and 
can be obtained already by simple models based on 
$U(1)$ charges or through 
sequential right--handed neutrino dominance \cite{King:2002nf}.
A correlation resulting from Eq.\ (\ref{eq:mnuNH1}) is 
\be \label{eq:corNH1} 
\meff = c_1 \, \sqrt{\dma} \, |U_{e3}|^2~\mbox{ with }~ |U_{e3}| = c_2\, R~,
\ee
where $c_{1,2}$ are functions of the order one parameters and naturally 
also of order one. With the parameters in the $\mu\tau$ block unspecified, 
the atmospheric neutrino mixing angle typically deviates sizable from its 
maximal value $\pi/4$, i.e., $\theta_{23} = \pi/4 - c_3 \, \sqrt{R}$.  

The discussion is also possible in the context of 
$\mu\tau$ symmetric mass matrices \cite{mutau,mutau2}. Since their 
predictions include maximal $\theta_{23}$ and zero 
$\theta_{13}$, one is interested in breaking schemes of the symmetry. 
Breaking the symmetry 
in the $e$ sector of $m_\nu$ leads to the same correlation 
\cite{rabimutau} as in Eq.\ (\ref{eq:corNH1}) but atmospheric 
neutrino mixing is much closer to $\pi/4$. 
On the other hand, breaking the $\mu\tau$ symmetry in the $\mu\tau$ sector  
of $m_\nu$ leads to \cite{rabimutau} 
\be 
\meff \simeq c_1 \, \sqrt{\dma} \, |U_{e3}|~, 
\ee
where $|U_{e3}|$ is now very small, $|U_{e3}| = c_2 \, R$, 
and the deviation from maximal atmospheric neutrino mixing is 
again sizable, i.e., of order $\sqrt{R}$. 
Another interesting breaking scenario of $\mu\tau$ symmetric mass matrices 
is when the $e\mu$ element is not identical to, but 
the complex conjugate of the $e\tau$ element \cite{mutaucp}. 
Then one finds that $U_{e3} \neq 0$ but Re$\, U_{e3}=0$, i.e., there is 
maximal $CP$ violation with $\delta = \pm \pi/2$. 
A special case of $\mu\tau$ symmetric mass matrices, 
for which similar statements will apply, is given when 
$\tan^2 \theta_{12}=0.5$, which then corresponds to the so--called 
tri--bimaximal mixing scheme \cite{tribimax}. 
``Bimaximal'' scenarios \cite{bimax} with their prediction 
$\tan^2 \theta_{12}=1$ typically require contributions 
from the charged leptons, see below.

Perturbing a zeroth order mass matrix that leads to zero $U_{e3}$ and maximal 
$\theta_{23}$ with a random matrix containing small entries of the same order 
$\epsilon$ \cite{adg}, 
will lead to a semi--hierarchical mass spectrum (i.e., $m_2 > m_1$ and not 
$m_2 \gg m_1$) with a somewhat larger 
\meff{} and also to sizable deviations both from $U_{e3}=0$ and 
$\theta_{23}=\pi/4$ \cite{adg}.

Other candidates studied frequently in the literature are minimal 
models in the sense of having zeros in $m_\nu$ \cite{2zeros} or a minimal 
number of parameters in a see--saw context \cite{minSS}. 

Table \ref{tab:corNH} summarizes our collection of the typical 
mass matrices for the normal mass hierarchy. Once we know that the 
neutrino mass spectrum is hierarchical, an information that could be 
provided by a negative search for \obb{} in future experiments and 
knowing that sgn$(\ma)=+1$, we can sort out 
the correct mass matrix when we have precise information on the 
oscillation parameters.\\

In case of the inverted hierarchy, 
summarized in Table \ref{tab:corIH}, the most stable candidate 
``theory'' corresponds to a mass matrix conserving the flavor charge 
$L_e - L_\mu - L_\tau$ \cite{lelmlt}. Such a 
matrix is given by 
\be \label{eq:lelmlt0}
m_\nu = m_0 
\left( 
\bad 
0 & \cos \theta & \sin \theta \\[0.3cm]
\cdot & 0 & 0 \\[0.3cm]
\cdot & \cdot & 0 
\ea 
\right)~, 
\ee
predicting one massless neutrino, 
zero $\theta_{13}$,  zero \meff, maximal {\it solar} 
neutrino mixing with $\dms = 0$ and $\theta_{23} = \theta$. 
Barring extreme fine--tuning, it is impossible to perturb the structure 
of Eq.\ (\ref{eq:lelmlt0}) such that the measured values of $\theta_{12}$ 
and \dms{} are both accommodated in accordance with the data. 
Typically, after breaking it will hold that 
$\tan^2 \theta_{12} \simeq 1 \pm R/2$ 
($\sin^2 \theta_{12} \simeq \frac{1}{2} \pm R/4$), whereas 
experimentally $\sin^2 \theta_{12} \simeq \frac{1}{2} - \sqrt{R}$ 
is required. 
An exception is a 
see--saw model in which the perturbations at high energy 
have the same order of magnitude as the terms allowed by 
the symmetry, see \cite{GL04}. In that example, 
$U_{e3}$ and the smallest mass state remain zero.   
Another way out is to take corrections from the charged lepton sector 
into account \cite{SP}. 
The matrix $U_\ell$, which diagonalizes 
$m_\ell m_\ell^\dagger$, is multiplied from the left to $U_\nu$, 
which is the matrix diagonalizing 
$m_\nu$ from Eq.\ (\ref{eq:lelmlt0}). The latter 
has to be perturbed in order to generate a non--zero \dms. 
Given the observation that the deviation from maximal solar neutrino mixing 
is determined by the Cabibbo angle, i.e., 
$\pi/4 - \theta_\odot \simeq \lambda \simeq 0.22 \simeq \theta_C$ 
\cite{QLC}, one assumes a CKM--like structure of $U_\ell$ 
\cite{STP}\footnote{Note that the so--called 
Quark--Lepton--Complementarity, i.e., 
the exact relation $\theta_{12} + \theta_C = \pi/4$ is a special case 
of this procedure.}.   
The typical result of the described Ansatz lies 
in a correlation between the solar neutrino mixing 
and $U_{e3}$, namely 
$ \sin^2 \theta_{12} \simeq \frac{1}{2} - \cos \phi \, \cot \theta_{23} 
\, |U_{e3}|$~, 
where $\phi$ is a phase appearing in $m_\nu$, which can be 
given by the Dirac phases as measurable in oscillation 
experiments. This depends however on the breaking in $m_\nu$.  
Nevertheless, both $|U_{e3}|$ and $\tan^2\theta_{12}$ are 
expected to be close to their current upper limits. 
Assuming minimal breaking in $m_\nu$ \cite{SP} 
one can show that there are remarkable correlations between the 
observables, such as 
$\meff \simeq \sqrt{\dma} \, 
\left|  \cos 2 \theta_{12} + 4 i \, J_{CP}/\sin^2 \theta_{23}
\right| $,  
where $J_{CP}$ is the Jarlskog invariant for $CP$ violating effects 
in neutrino oscillations, which is proportional to $|U_{e3}|/4$. 
%, one 
%can see from this relation that the effective mass has 
%sizable dependence on $|U_{e3}|$. Both terms in the expression for \meff{}  
%can be of the same order of magnitude, namely ${\cal O}(0.1)$. 
%The Dirac phase and the Majorana phases are in leading order identical. 
This represents one of the few examples where the $CP$ phases are part of 
the predicted correlations, depending however on the details of the model. 
The correlation $\tan^2 \theta_{12} \simeq 1 - 4~|U_{e3}|$ is however 
independent of the breaking in $m_\nu$ 
and relies only on bi--large $U_\nu$ \cite{STP}. 
Hence, the two discussed possibilities incorporating $L_e - L_\mu - L_\tau$ 
predict values for \meff{} of similar size but predict either zero 
or large $U_{e3}$. 

There is another zeroth order scheme for the inverted hierarchy, namely 
\be \label{eq:IH2}
m_\nu = m_0 
\left( 
\bad 
1 & 0 & 0 \\[0.3cm]
\cdot & 1/2 & 1/2 \\[0.3cm]
\cdot & \cdot & 1/2 
\ea 
\right)~. 
\ee
Zero $U_{e3}$ and maximal $\theta_{23}$ are predicted and the 
two leading mass states have the same sign. One can 
perturb this structure \cite{adg} and the result is that both 
$U_{e3}$ and $\theta_{23} - \pi/4$ are very small quantities, but 
the effective mass is larger than in case of matrices based on 
the conservation of $L_e - L_\mu - L_\tau$. 

The dependence of \meff{} 
on $U_{e3}$ goes from none to sizable to little in the three 
cases discussed, cf.\ with Table \ref{tab:corIH}. So, if we know 
that the mass ordering is inverted, i.e., sgn$(\ma)=-1$ and the 
effective mass is of order $\sqrt{\dma}$, then we basically only have to 
distinguish between $L_e - L_\mu - L_\tau$ from Eq.\ (\ref{eq:lelmlt0}) 
and the matrix given in Eq.\ (\ref{eq:IH2}). This can be done 
by knowing how to close \meff{} is to $\sqrt{\dma}$. More precision data 
on $U_{e3}$ and $\theta_{23}$ will fix the details of the model.\\

Now we turn to quasi--degenerate neutrinos. 
Usually, in such scenarios 
both the normal and the inverted mass ordering can be accommodated. 
Typical examples are given in Table \ref{tab:corQD}. 
There are matrices with two zero entries which are compatible 
with quasi--degenerate neutrinos and which predict an 
interesting interplay of observables \cite{2zeros}. Anarchical 
mass matrices \cite{UK} do in general not 
generate extreme mixing angles and therefore 
typically result in large deviations from zero $U_{e3}$ and maximal 
$\theta_{23}$, i.e., of order $R$.

Possibilities to obtain QD neutrinos are models based on 
flavor democracy \cite{demo}, of which we quote in Table \ref{tab:corQD} 
one interesting recent and typical example \cite{demozzx}. 
In such models both $U_\nu$ and $U_\ell$ 
are required in order to reproduce the neutrino data, 
resulting in a dependence of the 
neutrino observables on the charged lepton masses. This makes the predictions 
also a function of the democracy breaking scenario, therefore somewhat 
model--dependent. However, the smallness of $|U_{e3}|$ in such models 
can be attributed to the small ratios of the charged lepton masses.

One can also upgrade a hierarchical mass spectrum 
(e.g.\ resulting from sequential dominance) to a quasi--degenerate 
one by adding a term proportional to the unit matrix 
(e.g.\ connected to a $SO(3)$ symmetry) to it \cite{anki}. 
The neutrino mixing angles display approximately 
the same behavior as in the NH case (i.e., as from Eq.\ (\ref{eq:mnuNH1})), 
however, the effective mass is close to the common mass scale, i.e., 
there is little cancellation in \meff. Moreover, the larger $m_0$ the 
smaller the phases \cite{anki}.

Some attention has recently been caught by a model based on the 
discrete symmetry $A_4$ \cite{A4}. Applying the most general SUSY threshold 
corrections to the resulting mass matrix leads to predictions such as 
very large atmospheric mixing, purely imaginary $U_{e3}$ with an 
absolute value of order $\sqrt{R}$ and 
$\meff=m_0$. Similarly, one can use a simple Abelian symmetry corresponding 
to the conservation of 
the flavor symmetry $L_\mu - L_\tau$ \cite{CR} and perturb the resulting mass 
matrix\footnote{Note that this matrix is a special case 
of $\mu\tau$ symmetry.} 
\be \label{eq:mnuQD1}
m_\nu = m_0 
\left( 
\bad 
\cos \theta  & 0 &  0 \\[0.3cm]
\cdot & 0 & \sin \theta  \\[0.3cm]
\cdot & \cdot & 0
\ea 
\right)~,
\ee
to which the $A_4$ model corresponds when $\theta = \pi/4$. 
The deviations from zero $U_{e3}$ and maximal 
$\theta_{23}$ are small and inverse proportional to $m_0^2$. 
Note that interestingly most of the models presented here have 
$\meff \simeq m_0$, i.e., the mass parameters hopefully measured in 
direct laboratory searches and in \onbb experiments should be almost 
identical.

Another zeroth order scheme for quasi--degenerate neutrinos 
corresponds to a matrix of the form \cite{QD1} 
\be \label{eq:mnuQD2}
m_\nu = m_0 
\left( 
\bad 
0 & -1/\sqrt{2} &  1/\sqrt{2} \\[0.3cm]
\cdot & (1+\eta)/2 & (1+\eta)/2 \\[0.3cm]
\cdot & \cdot & (1+\eta)/2 
\ea 
\right)~.
\ee
Exact bimaximal mixing is predicted and 
breaking of the matrix is also required to generate 
splittings between the mass states.  
Similar statements as for the matrix Eq.\ (\ref{eq:lelmlt0}) hold, i.e., 
without extreme fine--tuning it is impossible to generate 
deviations from maximal solar neutrino mixing sizable enough not to 
be in conflict with the data. Hence, the contribution from the 
charged lepton sector are required, which will lead to similar correlations 
as for $L_e - L_\mu - L_\tau$ discussed above, such as 
$\tan^2 \theta_{12} \simeq 1 - 4~|U_{e3}|$. The correlation between the other 
observables depends strongly on the breaking. Since however 
the $ee$ element is zero in Eq.\ (\ref{eq:mnuQD2}), the effective mass 
is expected to be small, i.e., $\meff \simeq m_0 \, \cos 2 \theta_{12}$.

Hence, the identification of the mass matrix in case of a QD spectrum 
can be performed when information also from direct kinematical or from 
cosmological measurements is available. 
%There seem to be more 
%possibilities which are currently allowed, however, 
With the expected future precision data the identification of the mass matrix 
should also be possible.\\

To sum up this Section, a complete determination of the all 
neutrino observables and consequent identification of correlations 
will be able to discriminate 
between the various possible models. At least some of the possible 
candidates will be ruled out. Note finally that the values of the 
two Majorana phases, whose determination turned out to be an extremely 
challenging task, is not really required in order to identify the mass matrix.

\section{\label{sec:concl}Conclusions}
We have analyzed some aspects of the connection between 
neutrinoless double beta decay,
the neutrino mass scale and neutrino oscillation parameters. 
In particular, we concentrated on the question whether the expected future 
precision on the oscillation parameters simplifies certain important 
consequences of a measurement of, or improved upper limit on, 
\obb. We first summarized the current situation of the determination of the 
nine physical parameters of the neutrino mass matrix and the prospects 
of improving our knowledge about them with future experiments. 
Then we analyzed how \obb{} can help in distinguishing the normal 
mass ordering from the inverted one. We included the nuclear 
matrix element uncertainty in the 
analysis and pointed attention to the fact that distinguishing NH from IH 
depends on the value of the smallest neutrino mass $m_{\rm sm}$. 
We analyzed this point and found that in principle values of 
$m_{\rm sm} = 0.01$ eV are allowed, where of course the 
issue of the \nme{} can significantly spoil this possibility.

Then we investigated inasmuch \obb{} can be used to set a limit on 
the neutrino mass scale $m_0$ in case of a quasi--degenerate spectrum. 
We argued how the information from oscillation data and from \obb{} can be 
separated. Current limits on $m_0$ are weaker than direct ones but can be 
significantly improved with future \obb--experiments and more 
precise knowledge of the oscillation parameters $\sss$ and $\sch$.

Next we studied the implications of a very small, 
or even vanishing effective mass. Still insisting that 
neutrinos are Majorana particles, one can put in this case 
interesting constraints on parameters, in particular on $\sss$ and the 
Majorana phases.

Finally, we tried to perform a scan through the literature and identified 
neutrino mass matrices, which tend to arise frequently in many different 
models. We listed their predictions and correlations and 
argued how future precision date of the 
oscillation parameters and information from \onbb can help to 
identify the true neutrino mass matrix or at least to sort out 
many unsuccessful ones. 

\vspace{0.5cm}
\begin{center}
{\bf Acknowledgments}
\end{center}
This work was supported by the ``Deutsche Forschungsgemeinschaft'' in the 
``Sonderforschungsbereich 375 f\"ur Astroteilchenphysik'' 
and under project number RO--2516/3--1 (W.R.) and PPARC grant 
number PPA/G/O/2002/00479 (S.C.).

%\newpage

\hspace{-1.7cm}
\begin{table}[h]\hspace{-1.7cm}
\begin{tabular}{|c|c|c|} \hline 
Matrix $m_\nu/m_0$ & comments & correlations \\ \hline \hline
$ 
\left(
\bad
a \, \epsilon^2 &  b \, \epsilon & d \, \epsilon \\[0.3cm]
\cdot & e & f \\[0.3cm]
\cdot & \cdot & g 
\ea
\right) $ 
& $\ba {\rm Ref.\ \cite{NH1}} \\ {\rm e.g., simple~} U(1), {\rm broken~}L_e \\ 
\mbox{sequantial dominance} \ea $   
& $ \ba \meff = c_1 \, \sqrt{\dma} \, |U_{e3}|^2 
\\ |U_{e3}| = c_2 \, \sqrt{R},~
\theta_{23} = \frac{\pi}{4} - c_3 \, \sqrt{R}
\\ {\rm normal~ hierarchy} \ea $\\ \hline
$ 
\left(
\bad
a \, \epsilon &  b \, \epsilon & d \, \epsilon \\[0.3cm]
\cdot & 1 + f \, \epsilon & 1 + g \, \epsilon\\[0.3cm]
\cdot & \cdot & 1 + h \, \epsilon
\ea
\right) $ 
& $\ba {\rm Ref.~} \cite{adg} \\ 
{\rm perturbed~} m_\nu^0 \\
{\rm broken~} L_e \ea$    
& $\ba \meff = \frac{\sqrt{\Delta m^2_{\rm atm}}}{2} \, ( 1 + c_1~|U_{e3}| ) 
\\ |U_{e3}| = c_2 \, \sqrt{R} ,~
\theta_{23} = \frac{\pi}{4} - c_3 \, \sqrt{R}
\\ {\rm normal~ordering~}m_2 > m_1 \ea $ \\ \hline
$ 
\left(
\bad
a \, \epsilon^{2}\ &  b \, \epsilon & d \, \epsilon \\[0.3cm]
\cdot & 1 + \epsilon & 1 \\[0.3cm]
\cdot & \cdot & 1 + \epsilon 
\ea
\right) $ 
& $\ba {\rm Ref.\ \cite{rabimutau}} \\ \mu\tau~{\rm symmetry} \\ 
{\rm broken~in~}e~{\rm sector} \ea $   
& $ \ba \meff = c_1 \, \sqrt{\dma} \, |U_{e3}|^{2} 
\\ |U_{e3}| = c_2 \, \sqrt{R} ,~
\theta_{23} = \frac{\pi}{4} - c_3 \, R
\\ {\rm normal~hierarchy} \ea$ \\ \hline
$ 
\left(
\bad
a \, \epsilon^{2} &  b \, \epsilon & b \, \epsilon \\[0.3cm]
\cdot & 1 + d\epsilon & 1 \\[0.3cm]
\cdot & \cdot & 1 + \epsilon 
\ea
\right) $ 
& $\ba {\rm Ref.\ \cite{rabimutau}} \\ \mu\tau~{\rm symmetry} \\ 
{\rm broken~in~}\mu\tau~{\rm sector} \ea $   
& $ \ba \meff = c_1 \, \sqrt{\dma} \, |U_{e3}| 
\\ |U_{e3}| = c_2 \, R,~
\theta_{23} = \frac{\pi}{4} - c_3 \, \sqrt{R}
\\ {\rm normal~hierarchy} \ea $
\\ \hline 
$ 
\left(
\bad
0 &  0 & \epsilon \\[0.3cm]
\cdot & a  & b \\[0.3cm]
\cdot & \cdot & d  
\ea
\right) $ 
& $\ba {\rm Ref.\ \cite{2zeros}} \\ 
\mbox{2 zeros} \\
\mbox{very similar to } m_{ee} = m_{e\tau}=0\ea $   
& $ \ba \meff = 0  
\\ |U_{e3}| = \sqrt{\frac{R}{\cos 2 \theta_{12}}} \, 
\frac{\sin 2 \theta_{12}}{2 \, \tan \theta_{23}}  
\\ {\rm normal~hierarchy} \ea $
\\ \hline 
\begin{small} $ 
\left(
\bad
a^2 \, \epsilon &  a \sqrt{1 - a^2} \, \epsilon & 0 \\[0.3cm]
\cdot & b^2 + (1 - a^2) \, \epsilon   & b \, \sqrt{1 - b^2} \\[0.3cm]
\cdot & \cdot & 1 - b^2
\ea
\right) $ 
\end{small}
& $\ba {\rm Ref.\ \cite{minSS}} \\  
\mbox{minimal see--saw }  \ea $   
& $ \ba \meff = \sqrt{\dms} \, \sin^2 \theta_{12}  
\\ |U_{e3}| = \sqrt{R}/2  \, \sin 2 \theta_{12} \, \tan \theta_{23}
\\ {\rm normal~hierarchy},~m_1=0 \ea $
\\ \hline 
\end{tabular}
\caption{\label{tab:corNH}
Typical and popular 
neutrino mass matrices predicting the normal mass hierarchy. 
Given are some comments about 
their origin and order of magnitude 
predictions for correlations between the observables. The 
parameter $\epsilon$ is usually small, i.e., mostly of 
order $\lambda \simeq 0.22$ 
and the parameters $a,b,d,e,f,c_{1,2,3}$ are of order one. 
%With 
%``perturbed $m_\nu^0$'' we denote a mass matrix $m_\nu^0$ which leads to 
%zero $U_{e3}$ and $\theta_{23} = \pi/4$, and is corrected by a matrix 
%containing only small entries, see text and \cite{adg}.
} \hspace{-1.7cm}
\begin{tabular}{|c|c|c|} \hline \hspace{-1.7cm}
Matrix $m_\nu/m_0$ & comments & correlations \\ \hline \hline
$ 
\left(
\bad
1 + a \, \epsilon &  b \, \epsilon &  d \, \epsilon \\[0.3cm]
\cdot & 1/2 + f \, \epsilon & 1/2 + g \, \epsilon\\[0.3cm]
\cdot & \cdot & 1/2 + h \, \epsilon
\ea
\right) $ 
& $\ba {\rm Ref.\ \cite{adg}} \\ {\rm perturbed~} m_\nu^0 \ea $   
& $ \ba \meff = \sqrt{\dma} \, ( 1 + c_1 \, |U_{e3}| )  
\\ |U_{e3}| = c_2 \, R,~
\theta_{23} = \frac{\pi}{4} - c_3 \, R
\\ {\rm inverted~hierarchy} \ea $
\\ \hline
$ 
\left(
\bad
0  & a &  b \\[0.3cm]
\cdot & \epsilon^2 & 0 \\[0.3cm]
\cdot & \cdot & 0 
\ea
\right) $ 
& $\ba {\rm Ref.~} \cite{SP} \\ 
{\rm broken~}L_e - L_\mu - L_\tau \\
{\rm and~}U_\ell \sim V_{\rm CKM}
\ea $   
& $ \ba 
\begin{small}
\meff = \sqrt{\dma} \, 
\left| \cos 2 \theta_{12} + 4i/\sin^2 \theta_{23} \, J_{CP} \right|  
\end{small}
\\ \tan^2 \theta_{12} = 1 - 4 \, \cos \delta \, \cot \theta_{23} \, |U_{e3}| 
\\ {\rm inverted~hierarchy} \ea $
\\ \hline
%$ 
%\left(
%\bad
%a & \sqrt{2}b \, \cos \theta & \sqrt{2} b\, \sin \theta  \\[0.3cm]
%\cdot & d (1 + \cos \theta) &  d \, \sin \theta\\[0.3cm]
%\cdot & \cdot & d (1 - \cos \theta )
%\ea
%\right) $ 
%& $\ba {\rm Ref.~} \cite{GL04} \\ 
%\mbox{see-saw }L_e - L_\mu - L_\tau \\
%{\rm {\it strongly}~broken~}
%\ea $   
%& $ \ba 
%\sqrt{\dma} \, \cos 2 \theta_{12} \le \meff \le \sqrt{\dma} 
%\\ U_{e3} = 0 ,~\theta_{23}~ \rm large 
%\\ {\rm inverted~hierarchy,}~m_3=0 \ea $
%\\ \hline
\end{tabular}
\caption{\label{tab:corIH}
Same as Table \ref{tab:corNH} for the inverted mass hierarchy. 
We have $J_{CP} = 
1/8 \, \sin 2 \theta_{12} \, \sin 2 \theta_{23} \, \sin 2 \theta_{13} 
\, \cos \theta_{13} \, \sin \delta$ with the neutrino oscillation $CP$ 
phase $\delta$. 
%The last row is an explicit model, 
The correlations of the last row depend on the breaking of the mass matrix. 
For more complicated breaking, the phase $\delta$ will be some combination 
of phases.} 
\end{table}
\vspace{-2cm}

\hspace{-1.7cm}
\begin{table}[h]\hspace{-1.7cm}
\begin{tabular}{|c|c|c|} \hline 
Matrix $m_\nu/m_0$ & comments & correlations \\ \hline \hline
$ 
\left(
\bad
1  &  0 &  0 \\[0.3cm]
\cdot & 1 & 0  \\[0.3cm]
\cdot & \cdot & 1 
\ea
\right) + \ba \rm sequential \\ \rm dominance\ea$ 
& $\ba {\rm Ref.\ \cite{anki}} \\ \mbox{type II see--saw} \\
\mbox{upgrade}\ea $   
& $ \ba 
\meff \simeq m_0 
\\  |U_{e3}| = c_1 \, \sqrt{R},~
\theta_{23} = \frac{\pi}{4} - c_2 \, \sqrt{R}
\\ \mbox{phases shrink with } m_0
\ea $\\ \hline
$ 
\left(
\bad
\delta  &  -1/\sqrt{2} & (1 - \epsilon)/\sqrt{2}  \\[0.3cm]
\cdot & (1 + \eta)/2 & (1 + \eta - \epsilon)/2   \\[0.3cm]
\cdot & \cdot & (1 + \eta - 2\epsilon)/2 
\ea
\right) $ 
& $\ba {\rm Ref.\ \cite{QD1}} \\ \epsilon \leftrightarrow 
\mbox{radiative} \\ \mbox{corrections}\ea $   
& $ \ba 
\meff \sim m_0\,  \cos 2 \theta_{12} \\
\tan^2 \theta_{12} \simeq 1 - 4~c_1 ~|U_{e3}|\\
\mbox{(charged leptons required)}
\ea $\\ \hline
$ 
\left(
\bad
1  &  0 &  0 \\[0.3cm]
\cdot & 0 & 1  \\[0.3cm]
\cdot & \cdot & 0 
\ea
\right) $ 
& $\ba {\rm Ref.\ \cite{A4}} \\ A_4 \mbox{ plus threshold} \\
\mbox{corrections}\ea $   
& $ \ba 
\meff = m_0 \ge 0.3~\rm eV
\\ {\rm Re}~ U_{e3} = 0,~|U_{e3}| = c_1 \, \sqrt{R}\\
\theta_{23} \simeq \pi/4 
\ea $\\ \hline
 $ 
\left(
\bad
1  &  0 &  0 \\[0.3cm]
\cdot & 0 & -1  \\[0.3cm]
\cdot & \cdot & 0 
\ea
\right) $ 
& $\ba {\rm Ref.\ \cite{CR}} \\ 
L_\mu - L_\tau \\
\mbox{ plus perturbations} \ea $   
& $ \ba 
\meff = m_0~(1/\sqrt{2} +  c_1 \, |U_{e3}|) 
\\ |U_{e3}| = c_2 \, \dma/m_0^2 \ls 0.1 \\
\theta_{23} = \pi/4 - c_3 \, |U_{e3}|
\ea $\\ \hline
$ 
\left(
\bad
a  &  \epsilon &  0 \\[0.3cm]
\cdot & 0 & b  \\[0.3cm]
\cdot & \cdot & d 
\ea
\right) $ 
& $\ba {\rm Ref.\ \cite{2zeros}} \mbox{ 2 zeros}
\\ \mbox{similar to } m_{e\mu} = m_{\tau\tau}=0 \\
\mbox{and } m_{e\mu} = m_{\mu\mu}=0 \\
\mbox{and }m_{e\tau} = m_{\tau\tau}=0\ea $   
& $ \ba 
\meff \simeq m_0 \simeq 
\sqrt{\left|\frac{\Delta m^2_{\rm atm} \, \tan^4 \theta_{23}}
{1 - \tan^4 \theta_{23} }\right|}
\\ 
R \simeq \frac{1 + \tan^2 \theta_{12}}{\tan \theta_{12}}
 \, \tan 2 \theta_{23} \, 
{\rm Re}~ U_{e3}\\
\Rightarrow \theta_{23} \neq \pi/4\mbox{ and }{\rm Re}~ U_{e3}\simeq 0
\ea $
\\ \hline
$ 
\left(
\bad
a  &  \epsilon^2 &  b\, \epsilon \\[0.3cm]
\cdot & 0 & d  \\[0.3cm]
\cdot & \cdot & 0 
\ea
\right) $ 
& $\ba {\rm Ref.\ \cite{2zeros}} \\ \mbox{2 zeros}\ea $   
& $ \ba 
\meff \simeq m_0 \simeq \sqrt{\dma} \cot 2\theta_{12} /\cos\delta 
\\ {\rm Re}~ U_{e3} \simeq \cot 2 \theta_{12} \, \cot 2 \theta_{23} 
\ea $
\\ \hline
$ 
r_\nu \left(
\bad
1  &  1 &  1 \\[0.3cm]
\cdot & 1 & 1  \\[0.3cm]
\cdot & \cdot & 1 
\ea
\right) + c_\nu \left(
\bad
1  &  0 &  0 \\[0.3cm]
\cdot & 1 & 0  \\[0.3cm]
\cdot & \cdot & 1 
\ea
\right) $ 
& $\ba {\rm Ref.\ \cite{demozzx}} 
\\ S(3)_L \times S(3)_R 
\\ {\rm democracy~} \ea $   
& $ \ba 
\meff \simeq m_0, \mbox{ requires }r_\nu \ll 1
\\ |U_{e3}| \simeq \sqrt{m_e/m_\mu},~ 
\theta_{23} \mbox{ close to bound}\\
\mbox{depends on $m_{e,\mu,\tau}$ and breaking}
\ea $\\ \hline
$ 
\left(
\bad
a  &  b &  d \\[0.3cm]
\cdot & e & f  \\[0.3cm]
\cdot & \cdot & g 
\ea
\right) $ 
& $\ba {\rm Ref.\ \cite{UK}} \\ {\rm anarchy~} \ea $   
& $ \ba 
|U_{e3}| \mbox{ close to upper bound},\\
\theta_{23} \mbox{ close to bound} \\
\mbox{extreme hierarchy unlikely}
\ea $
\\ \hline
\end{tabular}
\caption{\label{tab:corQD}Same as Table \ref{tab:corNH} for the 
quasi--degenerate mass spectrum. The quantity measurable in direct mass 
measurements is denoted $m_0$.} 
\end{table}

%%%%%%%%%%%%%%%%%%%%%%%%%%%%%%%%%%%%%%%%%

\pagestyle{empty}
%%%%%%%%%%%%%%%%%%%%%%%%%%%%%%%%%%%%%%%%%
%%%%%%%%%%%%%%%%%%%%
\newpage

%%%%%%%%%%%%%%%%%%%%%%%%%%%%%%%%%%%%%%%%$$$$$$
%%%%%%%%%%%%%%%%%%%%
%%%%%%%%%%%%%%%%%%%%%%%%%%%%%%%%%%%%%%%%%%%%%%%%%%%%%%%%%%%
%%%%%%%%%%%%%%%%%%%%
%%%%%%%%%%%%%%%%%%%%%%%%%%%%%%%%%%%%%%%%%%%%%%%%%%%%%%%%%%%

%%%%%%%%%%%%%%%%%%%%
%%%%%%%%%%%%%%%%%%%%%%%%%%%%%%%%%%%%%%%%%%%%%%%%%%%%%%%%%%%

\end{document}